\documentclass[10pt]{iopart}

\usepackage[british]{babel}

\usepackage{mathtools} 
\usepackage{booktabs} 
\usepackage{tikz} 

\usepackage[utf8]{inputenc}
\usepackage{graphicx}
\usepackage{textcomp}
\usepackage{xcolor}
\usepackage{multirow}
\usepackage{amsfonts}
\usepackage{array}
\usepackage{tabularx}
\usepackage{caption}
\usepackage{subcaption}
\usepackage{hyperref}
\usepackage[inkscapelatex=false]{svg}

\usepackage{neuralnetwork}
\usepackage{siunitx}

\usepackage{algpseudocode}

\usepackage{tikz}
\usetikzlibrary{positioning}

\usepackage{tikz-imagelabels}

\usetikzlibrary{arrows.meta,shapes.arrows,chains,decorations.pathreplacing}
\usetikzlibrary{shapes.geometric}

\imagelabelset{
coarse grid color = red,
fine grid color = gray,
image label font = \sffamily\bfseries\small,
image label distance = 2mm,
image label back = white,
image label text = black,
coordinate label font = \sffamily\scriptsize, coordinate label distance = 2mm,
coordinate label back = white,
coordinate label text = black,
annotation font = \normalfont\small,
arrow distance = 1.5mm,
border thickness = 0.6pt,
arrow thickness = 0.4pt,
tip size = 1.2mm,
outer dist = 0.5cm,
}

\newlength\myht
\tikzset{%
  MyStyle/.style={draw, text width=25pt, text height=10pt, text centered,minimum height=\myht+2*2*1mm)},
  myarrow/.style={shape=single arrow, rotate=90, inner sep=5pt, outer sep=0pt, single arrow head extend=0pt, minimum height=7.5pt, text width=0pt, draw=blue!50, fill=blue!25}
}
\tikzset{database/.style={cylinder,aspect=0.5,draw,rotate=90,path picture={
\draw (path picture bounding box.160) to[out=180,in=180] (path picture bounding
box.20);
\draw (path picture bounding box.200) to[out=180,in=180] (path picture bounding
box.340);
}}}

\tikzset{%
  every neuron/.style={
    circle,
    draw,
    minimum size=0.8cm
  },
  neuron missing/.style={
    draw=none, 
    scale=4,
    text height=0.333cm,
    execute at begin node=\color{black}$\vdots$
  },
}

\definecolor{pxdark0}{HTML}{2E91E5}
\definecolor{pxdark1}{HTML}{E15F99}
\definecolor{pxdark2}{HTML}{1CA71C}
\definecolor{pxdark3}{HTML}{FB0D0D}
\definecolor{pxdark4}{HTML}{DA16FF}
\definecolor{pxdark5}{HTML}{222A2A}
\definecolor{pxdark6}{HTML}{B68100}
\definecolor{pxdark7}{HTML}{750D86}
\definecolor{pxdark8}{HTML}{EB663B}
\definecolor{pxdark9}{HTML}{511CFB}
\definecolor{pxdark10}{HTML}{00A08B}

\usepackage{marginnote}

\begin{document}

\title[Surrogate collisional radiative models from deep generative autoencoders]{Data-driven plasma modelling: Surrogate collisional radiative models of fluorocarbon plasmas from deep generative autoencoders}

\author{G A Daly$^{1,2*}$, J E Fieldsend$^1$, G Hassall$^{2}$ and G R Tabor$^1$}

\address{$^1$Faculty of Environment, Science and Economy, University of Exeter, North Park Road, Exeter, UK, EX4 4QF}
\address{$^2$Oxford Instruments Plasma Technology, North End, Yatton, UK, BS49 4AP}
\ead{*gd351@exeter.ac.uk}

\begin{abstract}


We have developed a deep generative model that can produce accurate optical emission spectra and colour images of an ICP plasma using only the applied coil power, electrode power, pressure and gas flows as inputs -- essentially an empirical surrogate collisional radiative model. An autoencoder was trained on a dataset of 812,500 image/spectra pairs in argon, oxygen, Ar/O\textsubscript{2}, CF\textsubscript{4}/O\textsubscript{2} and SF\textsubscript{6}/O\textsubscript{2} plasmas in an industrial plasma etch tool, taken across the entire operating space of the tool. The autoencoder learns to encode the input data into a compressed latent representation and then decode it back to a reconstruction of the data. We learn to map the plasma tool's inputs to the latent space and use the decoder to create a generative model. The model is very fast, taking just over 10 s to generate 10,000 measurements on a single GPU. This type of model can become a building block for a wide range of experiments and simulations. To aid this, we have released the underlying dataset of 812,500 image/spectra pairs used to train the model, the trained models and the model code for the community to accelerate the development and use of this exciting area of deep learning. Anyone can try the model, for free, on Google Colab. 

\end{abstract}

%
\maketitle
%
\ioptwocol

\section{Introduction}\label{sec:intro}

Generative models are a type of deep learning model that can produce new, unseen samples when trained on un-labeled data. These types of models have not been used previously in the field of low-temperature plasmas, but have been used to great effect in generating text, images and 3D models. They can offer many benefits by creating synthetic data for modelling and experiment design, replacing parts of computational models with fast surrogate models and providing a foundation for models that predict expensive and difficult to measure parameters from simpler diagnostics.



\subsection{Background}

Synthetic data can be an extremely useful resource in plasma physics for developing experiments, understanding diagnostics and training models and controllers for plasma applications. Synthetic data tools have been used in fusion \cite{shiSyntheticDiagnosticsPlatform2016, jacobsenInversionMethodsFastion2016, dalsaniaApplicationMachineLearning2021, juvenTemperatureEstimationFusion2022} and laser plasmas \cite{siminosModelingUltrafastShadowgraphy2016, crillySyntheticNuclearDiagnostics2018, milderImpactNonMaxwellianElectron2019, lewisDeeplearningenabledBayesianInference2021, rodimkovMLBasedDiagnosticsLaser2021} to aid simulations, experiment design and for training Machine Learning (ML) and Deep Learning (DL) models. However, such approaches have been used less frequently in low-temperature plasmas \cite{boffardOpticalEmissionMeasurements2010, liuOHConcentrationTemperature2022, gergsEfficientPlasmasurfaceInteraction2022}.

Methods for generating synthetic data, used in plasma physics, can be split into three main groups -- generating synthetic sensor data from simulation or analytic models \cite{shiSyntheticDiagnosticsPlatform2016, siminosModelingUltrafastShadowgraphy2016, crillySyntheticNuclearDiagnostics2018, milderImpactNonMaxwellianElectron2019, lewisDeeplearningenabledBayesianInference2021, rodimkovMLBasedDiagnosticsLaser2021, juvenTemperatureEstimationFusion2022, gergsEfficientPlasmasurfaceInteraction2022}, inverting analytic methods for extracting parameters from sensor data \cite{boffardOpticalEmissionMeasurements2010, jacobsenInversionMethodsFastion2016, liuOHConcentrationTemperature2022} and augmenting existing experimental data to create new data \cite{dalsaniaApplicationMachineLearning2021}. However, DL generative models have not been used for synthetic data generation in plasma physics. This approach uses DL models, such as autoencoders (AE), generative adversarial networks, diffusion models or transformers as a generative model that can create new synthetic data (see \cite{bond-taylorDeepGenerativeModelling2022} for a recent review of the area). Outside the field these approaches have been used for improving medical image classification \cite{frid-adarSyntheticDataAugmentation2018}, drug design \cite{chengMolecularDesignDrug2021}, chemical reaction discovery \cite{tempkeAutonomousDesignNew2022}, cyber security \cite{lopez-martinVariationalDataGenerative2019}, music generation \cite{choiEncodingMusicalStyle2020} and image generation \cite{rameshHierarchicalTextConditionalImage2022}, and many other applications besides.

Deep learning approaches have had many successes in the field, applied to controlling atmospheric pressure plasma jets \cite{witmanSimtorealTransferReinforcement2019, mesbahMachineLearningModeling2019}, a fast replacement for computed tomography for tokamak radiation profiles \cite{ferreiraDeepLearningPlasma2020}, predicting electron energy distribution functions from optical emission spectra (OES) \cite{shojaeiApplicationMachineLearning2021} and creating surrogate models of neutral beam injection \cite{boyerRealtimeCapableModeling2019}, sputtering processes \cite{gergsEfficientPlasmasurfaceInteraction2022} and plasma etching \cite{maggipintoDeepVMDeepLearningbased2019}.

In this work we demonstrate how deep autoencoders can be used to generate synthetic sensor data from large amounts of unlabelled experimental data. We show how to train a deep autoencoder on unlabelled data and then how to train a model to learn to `map' from an input space of physical variables into the latent space of the autoencoder to produce a generative model.

In the context of the literature on deep learning, there has been a great deal of interest in developing generative models for some time, such as variational autoencoders (VAE) \cite{kingmaAutoEncodingVariationalBayes2014}, generative adversarial models \cite{goodfellowGenerativeAdversarialNetworks2014} and diffusion models \cite{hoDenoisingDiffusionProbabilistic2020}. Earlier work focused on developing models that were capable of generating good outputs through random sampling, more recent work has focused on how to guide generative models to produce desired generative outputs. This can be referred to as learning a prompt for generative output or a map to a latent space. Recent examples include generating music \cite{choiEncodingMusicalStyle2020, dhariwalJukeboxGenerativeModel2020}, transforming facial expressions  \cite{nitzanFaceIdentityDisentanglement2020} generating energy angle distributions in sputtering processes \cite{gergsEfficientPlasmasurfaceInteraction2022}, and new high quality image generation from prompt models such as DALL·E 2, Parti and Stable Diffusion \cite{rameshHierarchicalTextConditionalImage2022, yuScalingAutoregressiveModels2022, rombachHighResolutionImageSynthesis2022}.

\subsection{Autoencoders}

\begin{figure}[t]
	\begin{center}
		\begin{tikzpicture}[x=1.0cm, y=1.5cm, >=stealth]

		\foreach \m/\l [count=\y] in {1,2,missing,3}
		\node [every neuron/.try, neuron \m/.try] (input-\m) at (0,2.5-\y) {};

		\foreach \m [count=\y] in {1,missing,2}
		\node [every neuron/.try, neuron \m/.try ] (hidden-\m) at (2,2.2-\y*1.15) {};

		\foreach \m [count=\y] in {1,2,missing,3}
		\node [every neuron/.try, neuron \m/.try ] (output-\m) at (4,2.5-\y) {};

		\foreach \l [count=\i] in {1,2,r}
		\draw [<-] (input-\i) -- ++(-1,0)
			node [above, midway] {$x_\l$};

		\foreach \l [count=\i] in {1,l}
		\node [above] at (hidden-\i.north) {$z_\l$};

		\foreach \l [count=\i] in {1,2,r}
		\draw [->] (output-\i) -- ++(1,0)
			node [above, midway] {$\hat{x}_\l$};

		\foreach \i in {1,...,3}
		\foreach \j in {1,...,2}
			\draw [->] (input-\i) -- (hidden-\j);

		\foreach \i in {1,...,2}
		\foreach \j in {1,...,3}
			\draw [->] (hidden-\i) -- (output-\j);

		\foreach \l [count=\x from 0] in {Input, Hidden, Output}
		\node [align=center, above] at (\x*2,2) {\l \\ layer};
		
		\foreach \l [count=\x from 0] in {$f(\cdot)$, $\mathbf{z}$, $g(\cdot)$} 
		\node [align=center, below] at (\x*2,-2) {\l};

		\end{tikzpicture}
		
		\end{center}
	\hspace{1cm}{\caption{\small Schematic diagram of a basic autocencoder.}\label{fig:aeschematic}}
	
\end{figure}
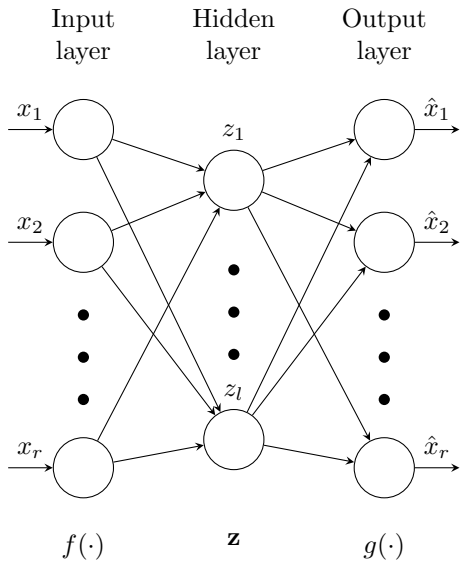

Autoencoders are an early type of neural network model that learns to copy its input at its output \cite{goodfellowDeepLearning2016}. Autoencoders consist of an encoder, $\mathbf{z} = f(\mathbf{x})$, that learns to map input data, $\mathbf{x}\in \mathbb{R}^r$, into a latent space ($\mathbf{z} \in \mathbb{R}^l$) and a decoder, $\hat{\mathbf{x}} =g(\mathbf{z})$ that learns to map the latent space representation back to the input \cite{goodfellowDeepLearning2016}, see figure \ref{fig:aeschematic}. The model is trained to minimise the reconstruction error between the input data and the reconstructed output. On the face of it this does not seem like a very useful network, but by making the latent space much smaller than the input data ($l<<r$), the network is forced to learn a low dimensional representation of the input data by learning relationships and patterns within the input data. 

VAEs are an extension of ordinary autoencoders, where an additional training objective, the Kullback–Leibler (KL) divergence, is added to guide the distribution of the latent space to follow a normal distribution with a diagonal covariance matrix, $\mathbf{z} = \mathcal{N} (\mathbf{z; 0, I})$. This gives VAEs a continuous latent space that can be easily sampled from to generate new samples. This has lead to VAEs being widely used in the field of generative modelling, however, they have had issues from their inception, as they are difficult to train and suffer from mode collapse \cite{vandenoordNeuralDiscreteRepresentation2017, heLaggingInferenceNetworks2019} and that the latent space does not always end up having the desired property of being a normal distribution \cite{kingmaImprovedVariationalInference2016}, such as in figure 4 of \cite{gergsEfficientPlasmasurfaceInteraction2022}.

Recent work in the field of generative modelling  has demonstrated that the VAE process can actually hamper the ability of the model to learn a useful representation through over-regularisation and that large autoencoders are good generative models, outperforming VAEs repeatedly \cite{ghoshVariationalDeterministicAutoencoders2019, ghoseBatchNormEntropic2020, dalyVariationalAutoencodersVariation2022}. In recent work, Autoencoders have been used to learn features for virtual metrology models from optical emission spectroscopy (OES) \cite{maggipintoConvolutionalAutoencoderApproach2018} and defect detection in semiconductor processing \cite{zhangAutomatedFaultDetection2020}. We use autoencoders in this work as they are easier and more predictable to train than VAEs, while providing equal or better performance as a generative model, making them more suitable for widespread use in scientific applications.

Our contributions in this work and the structure of the paper are laid out as follows. In section \ref{sec:intro} we provide a background to synthetic data generation, deep generative models and how it has been applied in other fields. In section \ref{experimentdesign} we describe how we created an experiment to gather 812,500 optical emission spectra and colour images in fluorocarbon plasmas in an industrial plasma etcher. In section \ref{aedesign} we describe how to build and train an autoencoder and how to train a small model to map physical tool inputs to the latent space and turn the decoder into a conditional generative model. In sections \ref{latentspaces} and \ref{geneval} we look at the structure of the latent space produced by the model for different sizes of latent space and the difficulty of evaluating generative models. In section \ref{synexp} we demonstrate using the generative model to carry out synthetic experiments looking at line ratios in Argon and Ar/O\textsubscript{2} plasmas covering 10,000 points varying power and pressure in seconds. We consider any limitations of the approach and future work, and detail the open source release of code and experimental results in sections \ref{limit} and \ref{foss}, followed by a conclusion to the work in section \ref{conc}. 

The data set we have gathered has been released under a creative commons license (CC BY-4.0) and can be used by anyone for academic purposes. The model's code and pre-trained models have been released as open source under the MIT License.

\section{Data collection and experimental design}\label{experimentdesign}

\begin{table*}[t!]
  \caption{\label{tab:setpoints}Dataset setpoints.}
  \footnotesize
  \lineup
  \begin{tabular*}{\textwidth}{@{}l*{15}{@{\extracolsep{0pt plus 12pt}}l}}
\br

 & Argon & Oxygen & Ar/O\textsubscript{2} & CF\textsubscript{4}/O\textsubscript{2} & SF\textsubscript{6}/O\textsubscript{2} \\
\mr
ICP / W & 480\textrightarrow 3000 & 600\textrightarrow3000 & 750\textrightarrow3000 & 600\textrightarrow3000 &  750\textrightarrow3000\\

Table / W & 0\textrightarrow 600 & 30\textrightarrow600 & 30\textrightarrow540 & 30\textrightarrow600 &  30\textrightarrow600 \\

Pressure / mT & 5\textrightarrow90 & 5\textrightarrow90 & 5\textrightarrow80 & 4\textrightarrow90 & 5\textrightarrow80\\

1st gas / sccm & 3.5\textrightarrow70 & 2.5\textrightarrow50 & 2.5\textrightarrow50 & 4.2\textrightarrow84 & 2.6\textrightarrow52\\

2nd gas / sccm &  &  & 2.5\textrightarrow50 & 2.5\textrightarrow50 &  2.5\textrightarrow50\\

Number of points, $n$ & 10,000 & 10,000 & 30,000 & 60,000 & 70,000\\

\br
\end{tabular*}\end{table*}\normalsize

A dataset of 812,500 optical emission spectra (OES) and RGB images of the bulk plasma above the wafer surface were gathered from an Oxford Instruments Plasma Technology PP 100 industrial plasma etcher with a Cobra300 cylindrical ICP source. Quartz windows were used for all optical diagnostics, for OES an Edmund Optics UV/VIS collimator (88-173) was used to collect light into a Thorlabs round to linear fibre bundle, consisting of seven \SI{200}{\micro\meter} solarisation resistant fibres. An Avantes ULS4096CL-EVO-RM 200-1100 nm spectrometer was used with a \SI{10}{\micro\meter} slit. Optical images were collected with a FLIR Blackfly 0.4 MP colour camera (BFS-U3-04S2M-CS) and a 6mm focal length lens (SV-0614V).

Data was collected across the entire operating region of the plasma source in argon, oxygen, Ar/O\textsubscript{2}, CF\textsubscript{4}/O\textsubscript{2} and SF\textsubscript{6}/O\textsubscript{2}. The experimental operating space consisted of the power delivered to the ICP source, the power to the table, the pressure in the chamber and the flow rate of one or two gases. The operating space varied for each gas due to differing lower limits on the minimum power and pressure to form a stable plasma or the requirement to keep the DC bias below 1kV. The operating space is summarised in table \ref{tab:setpoints}.

\begin{table*}[t!]
  \caption{\label{tab:actualpoints}Raw measured points.}
  \footnotesize
  \lineup
  \begin{tabular*}{\textwidth}{@{}l*{15}{@{\extracolsep{0pt plus 12pt}}l}}
\br

 & Argon & Oxygen & Ar/O\textsubscript{2} & CF\textsubscript{4}/O\textsubscript{2} & SF\textsubscript{6}/O\textsubscript{2}\\
\mr
ICP / W & 0\textrightarrow 2997 & 0\textrightarrow2996 & 0\textrightarrow2997 & 68.1\textrightarrow2996 & 0\textrightarrow2996\\
ICP 0.1\%\textrightarrow99.9\% & 464\textrightarrow2985 & 544\textrightarrow2988 & 72.2\textrightarrow2988 & 224\textrightarrow2988 &  595\textrightarrow2988 \\
Table / W & 0\textrightarrow 613 & 0\textrightarrow604 & 0\textrightarrow544 & 2.75\textrightarrow614 & 0\textrightarrow545\\
Table 0.1\%\textrightarrow99.9\% & 0.2\textrightarrow597 & 19.7\textrightarrow598 & 8.9\textrightarrow537 & 83.2\textrightarrow597 &  6.6\textrightarrow535\\

Pressure / mT & 5\textrightarrow92 & 5\textrightarrow91 & 2.8\textrightarrow91 & 3.8\textrightarrow85.8 & 4.3\textrightarrow82.1\\

1st gas / sccm & 3.5\textrightarrow70 & 0.1\textrightarrow50 & 2.8\textrightarrow70 & 4.2\textrightarrow84 & 0\textrightarrow52\\

2nd gas / sccm &  &  & 2.5\textrightarrow50 & 2.5\textrightarrow50 &  0\textrightarrow50\\

Number of points & 50,000 & 50,000 & 150,000 & 225,000 & 337,500\\

\br
\end{tabular*}\end{table*}\normalsize

Our aim was to make measurements at sample points across the operating space and gather the most amount of information within a fixed budget of samples. Naively, we could have used a grid search, however, a 10 point grid across 5 dimensions would require 100,000 points with very poor space filling, i.e there would be only 10 unique values in each dimension. The next simplest approach would be to sample randomly, for large numbers of samples -- this is quite likely to fill the parameter space, but there is no guarantee on how efficiently we can fill the operating space. The efficiency of filling a space and how well the points are separated can be measured by the discrepancy of the entire set, in particular, we use the L2 discrepancy to measure this \cite{jaeckelMonteCarloMethods2002, morokoffQuasiRandomSequencesTheir1994}. 

Quasi-random sequences offer a very effective way to generate sets of sample points that offer some guarantees on efficiency of filling a parameter space while still providing enough random spread to cover the interactions of many variables \cite{jaeckelMonteCarloMethods2002, morokoffQuasiRandomSequencesTheir1994}, i.e. they have a low discrepancy. Two of the most common quasi-random sequences are Latin Hypercube Sampling (LHS) and Sobol sequences, both have the properties that we desire, but Sobol sequences have an advantage the you can generate further elements of the sequence, using the same random seed. This is important if you need to extend your dataset at a later time point. There is no guarantee that the combination of two LHS sets does not have a higher discrepancy than one generated with the combined number of data points and you cannot truncate or randomly sample from a large LHS and maintain the low discrepancy. However, with a Sobol sequence you have a guarantee that the extension to your dataset has the same discrepancy as if you had started by generating the sequence of that length \cite{jaeckelMonteCarloMethods2002, sobolDistributionPointsCube1967a}.

\Table{\label{tab:l2discrepancy}L2 discrepancy of different sampling methods in 5 dimensions (lower is better, bold is best).}
\br
        No. points & Grid        & Random      & Sobol       \\ \hline
        $10^3$       & $1.14\times10^{-1}$   & $1.54\times10^{-3}$ & $\mathbf{2.52\times10^{-5}}$ \\
        $10^4$      & $2.87\times10^{-2}$ & $1.08\times10^{-4}$ & $\mathbf{1.83\times10^{-7}}$ \\
        $10^5$        & $1.38\times10^{-2}$  & $1.80\times10^{-5}$ & $\mathbf{4.63\times10^{-9}}$ \\
        $10^6$        & $5.09\times10^{-3}$  & $1.28\times10^{-6}$ & $\mathbf{1.03\times10^{-10}}$ \\

\br
\endTable

\begin{figure*}[t]
    \centering
    \includegraphics[width=1.0\textwidth, trim={2.5cm 2.5cm 2.5cm 3.0cm}, clip]{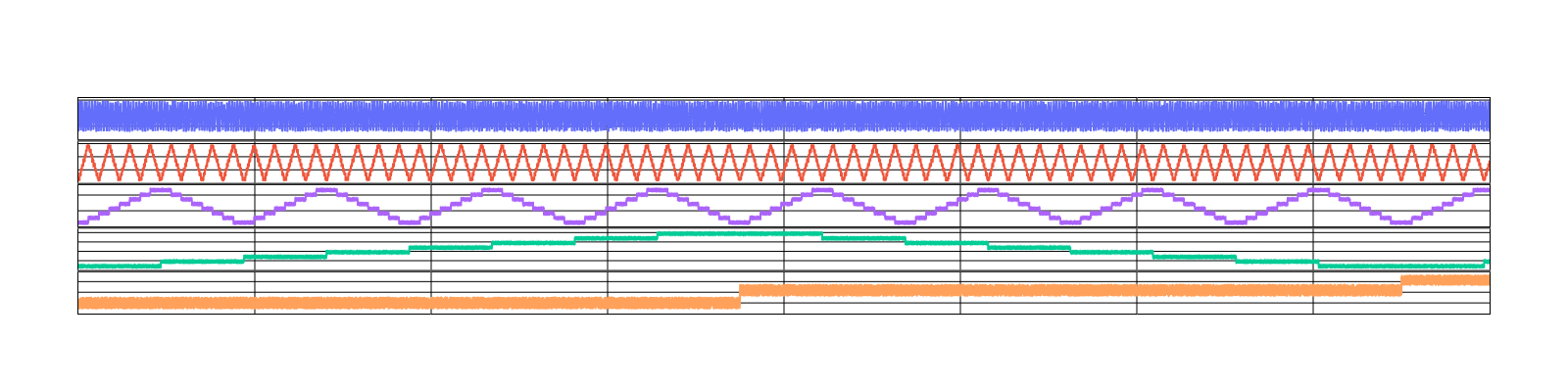}
    \caption{Ordered sweep from Sobol sequence for CF\textsubscript{4}/O\textsubscript{2}, from top to bottom - IPC power, Table power, O\textsubscript{2} flow, CF\textsubscript{4} flow, pressure.}
    \label{fig:sobol_sweep}
\end{figure*}

Using a Sobol sequence, we generated 10,000 points each for argon and oxygen, 30,000 points for Ar/O\textsubscript{2} and 60,000 for CF\textsubscript{4}/O\textsubscript{2} and 70,000 for SF\textsubscript{6}/O\textsubscript{2}. To actually cover the entire sequence in our experiment, we sorted each sequence such that pressure followed a relatively flat ramp over the whole range and other variables followed a triangle wave shape of increasing speed, as shown in figure \ref{fig:sobol_sweep}. This enabled us to maintain tool stability between sample points and reduced the settling time between setpoint changes. Setpoints were changed every 5 seconds and a optical image and OES were taken every second starting at the beginning of the setpoint change. A plain, un-patterned, silicon wafer was clamped to the table at all times and the process was only stopped to replace the wafer when it had become too thin from etching.

The dataset consists of 5 image spectra pairs, $[i_{n,0}, \ldots, i_{n,4}]$, $[s_{n,0}, \ldots, s_{n,4}]$ and setpoint readbacks from the tool $[t_{n,0}, \ldots, t_{n,4}]$, taken at each setpoint $[P_0,\ldots,P_n]$ for each gas mixture. The setpoint readbacks consist of the net power (forward-reflected) on the ICP coil and table, pressure in the chamber, gas flow from each mass flow controller and DC bias at the table.


The experimental points sampled did not perfectly align with our planned sweeps; some areas had unstable plasmas, could not sustain a plasma or exceeded parts of the tool's operational envelope, such as pressure control. The measured data is summarised in table \ref{tab:actualpoints}, all of the runs have a small portion of results with momentary high reflected power, but not for long enough to cause the plasma to extinguish. In CF\textsubscript{4}/O\textsubscript{2} plasma the high pressure region above 70 mT was unstable due to a combination of reduced plasma stability and limited control margin of the pressure controller and the sweeps were not continued above this pressure. In SF\textsubscript{6}/O\textsubscript{2}, the minimum power required to sustain a plasma increased with pressure and so the sequence was extended to 70,000 points and the minimum ICP power raised to 1500 W above 40 mT to yield more measurement points. The experiment yielded a total of 812,500 image spectra pairs, at 162,500 unique setpoints in the operational space of the tool.

\begin{figure}[t]
    \centering
    \includegraphics[width=0.25\textwidth, trim={4cm 3cm 1.5cm 3cm}, clip]{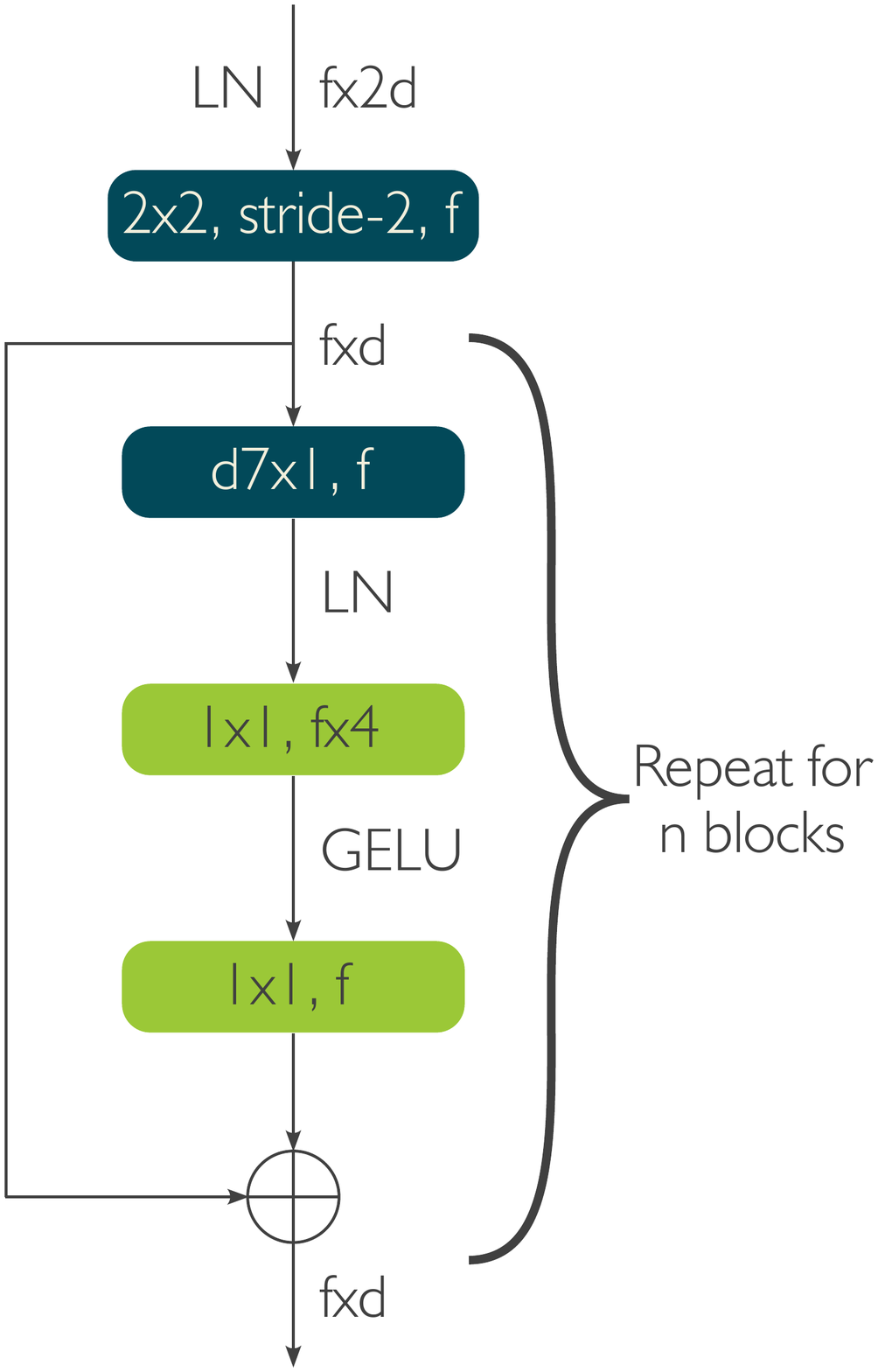}
    \caption{1D Convnext basic block with layer normalisation (LN) and Gaussian Error Linear Unit (GELU) activations.}
    \label{fig:convnext_block}
\end{figure}


\begin{figure*}[t!]
    \centering
    \begin{annotationimage}{width=1.0\linewidth, trim={0 0 0 2.4cm}, clip}{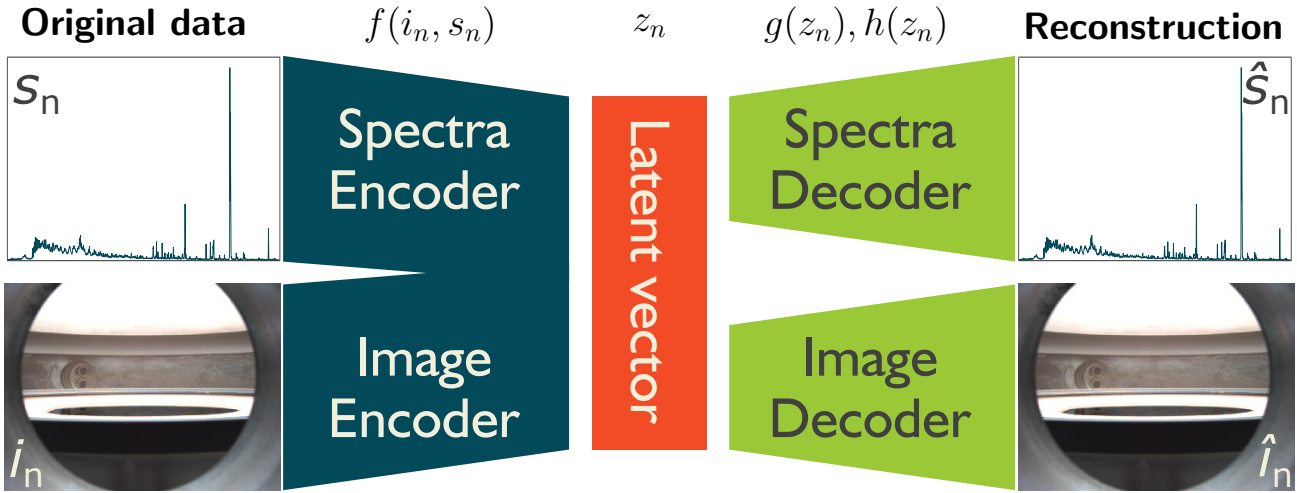}
    \draw[coordinate label = {\Large{\textbf{Original data}} at (0.10,1.05)}];
    \draw[coordinate label = {\Large{$f(i_n, s_n)$} at (0.33,1.05)}];
    \draw[coordinate label = {\Large{$z_n$} at (0.5,1.05)}];
    \draw[coordinate label = {\Large{$g(z_n),h(z_n)$} at (0.66,1.05)}];
    \draw[coordinate label = {\Large{\textbf{Reconstruction}} at (0.89,1.05)}];
    \end{annotationimage}
    \caption{Autoencoder architecture block diagram.}
    \label{fig:ae_block_diagram}
\end{figure*}

The data was split into train, validation and test sets with a 80/10/10 split. However, since we hold and take 5 measurements at each set point, naively randomly splitting the data would result in leakage from the test data into the train split, i.e. some measurements at a single setpoint would be present in each split. To avoid this, the data is kept together in blocks of 5 and the blocks are randomly assigned to the three sets. The spectra are processed by subtracting the average of the counts at the dark pixels from each spectra and removing the data from pixels outside the calibrated range of the spectrometer, this leaves 3072 pixels covering 200-1100 nm. The intensity of each spectra is min-max scaled to between 0 and 1 and a 5 pixel wide Hann window \cite{blackmanMeasurementPowerSpectra1958} is used to smooth out noise in the spectra. The camera produces a 720x540 pixel image with an RGGB Bayer mask, rather than perform standard Bayer interpolation to produce a 720x540 colour image, we treat the camera like a hyperspectral camera with very poor spectral resolution. We take all the red and blue pixels and one of the green pixels to form three 360x270 images. These are cropped to the central area of the image, resized and stacked to produce a 128x96x3 image. The pixel intensities are well controlled by the camera's autoexposure algorithm and are all clustered around a 50\% grey value, requiring no further normalisation. The camera ADC is set to a 10-bit resolution and values are stored as 16-bit integers, all images are divided by $2^{16}$ to rescale their pixel intensities between 0 and 1. The values from the tool's setpoint readbacks are all in the range of 0-10 V or 0-5 V and are simply divided by 10 to rescale them between 0 and 1.

This process of the rescaling and normalisation of inputs is a particularly important step in preparing data for training in any machine learning approach. It speeds up and stabilises convergence in training the model \cite{lecunEfficientBackProp1998, ioffeBatchNormalizationAccelerating2015}, as gradients in the model will be within expected bounds for the optimiser and the inputs are within the expected bounds of activation functions, such as sigmoid and ReLU.




\section{Building deep generative autoencoders for synthetic data generation}\label{aedesign}


Our model architecture is based on ConvNeXt, a state of the art convolutional neural network architecture \cite{liuConvNet2020s2022}. We use the base ConvNeXt blocks and stem, with 1D or 2D convolutions for OES or images to form our image and spectra encoding branches, the basic block is shown in figure \ref{fig:convnext_block}. Each branch consists of four stages with (2, 2, 6, 2) blocks and (64, 128, 256, 512) filters, at the beginning of each stage a convolutional downscaling halves the spatial dimensions of the image or spectra. At the end of the last stage a global average pooling layer reduces all of the spatial dimensions and produces a single tensor with the size of the last set of filters and this is followed by two densely connected neural network layers of 1024 neurons and the chosen size of our latent space. The latent output of each branch is then summed together producing a tensor with the length of the latent space dimension and finishes in a dense layer with $\mathbf{z}$ neurons with a linear activation function. This is our latent representation of the input data and can be the combination of any number of input branches. The model was trained on different sized latent spaces, $l = [4,16,32,64]$, to demonstrate the effect the size of the latent space has on the model.

In this work we have only used two branches, both based on convolutional networks, but any number of branches can be used with any kind of network architecture encoding some input data. The decoder is simply the reverse of the encoder and finishes in a 1D or 2D convolution that reconstructs the input.

\Table{\label{tab:trainingsetttings}Settings for autoencoder model training and fine-tuning.}
\br

config & Training & Finetune \\
\mr
optimiser & Adam & Adam \\
epochs & 100 & 100 \\
base learning rate & $2.5\times10^{-4}$ & $1\times10^{-4}$ \\
learning rate schedule & cosine decay & cosine decay \\
warmup epochs & 8 & 8 \\
warmup schedule & linear & linear \\
batch size & 2048 & 2048 \\
blocks & 2,2,6,2 & 2,2,6,2 \\
filters (f) & 64, 128, 256, 512 & 64, 128, 256, 512 \\
\br
\endTable

The encoder learns a function to project the input image and spectra $i_n, s_n$ pair into a latent space, $z_n = f(i_n, s_n)$, each decoder branch then learns a function to project the latent space vector back into the real diagnostic space, $\hat{i_n} = g(z_n)$, $\hat{s_n} = h(z_n)$, this overall structure is shown in figure \ref{fig:ae_block_diagram}. The loss is a reconstruction loss between input, $i_n, s_n$, and reconstructions, $\hat{i_n}, \hat{s_n}$. This loss can be weighted to favour one input over another to embed prior assumptions about the relative importance of each diagnostic.

The model is trained with the Adam optimiser \cite{kingmaAdamMethodStochastic2015}, using a cosine decay learning rate schedule \cite{loshchilovSGDRStochasticGradient2017} with a linear warmup, and Mean-Squared Error (MSE) as the loss, using Keras \cite{cholletKeras2015}/Tensorflow \cite{abadiTensorFlowLargeScaleMachine}. Full details of the training and fine-tuning settings are in table \ref{tab:trainingsetttings}. The model was trained on 4 Nvidia A100 GPUs for 100 epochs, taking roughly 20.5 hours to train.

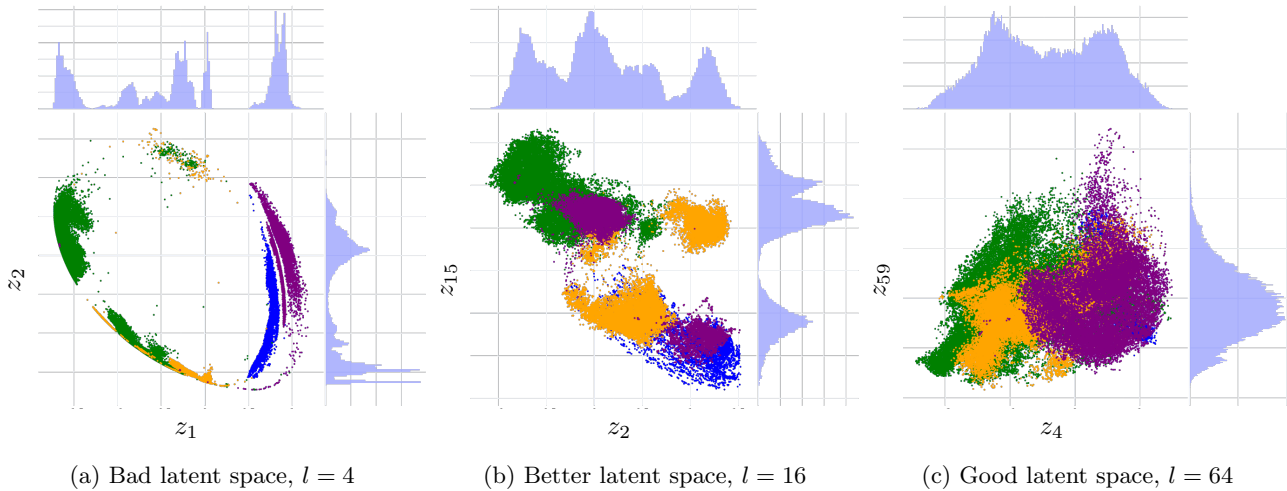
\begin{figure*}[htbp!]
    \centering
    \begin{subfigure}[b]{0.329\textwidth}
        \centering
        \begin{annotationimage}{width=1.0\linewidth, trim={3.05cm 2.65cm 0 0}, clip}{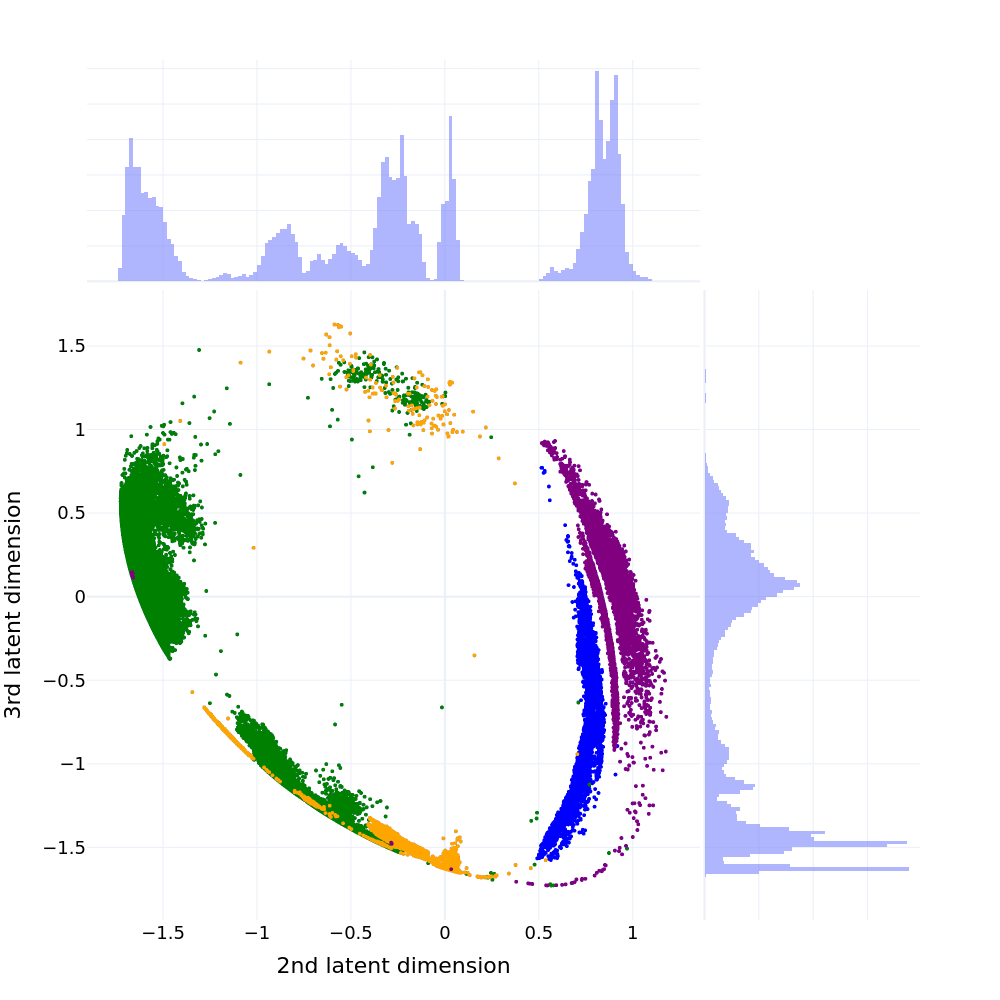}
            \draw[coordinate label = {\normalsize{$z_1$} at (0.35,-0.05)}];
            \draw[coordinate label = {\rotatebox{90}{\normalsize{$z_2$}} at (-0.05,0.3)}];
        \end{annotationimage}
        \caption{Bad latent space, $l =4$}
        \label{fig:bad_latent}
     \end{subfigure}
     \hfill
    \begin{subfigure}[b]{0.329\textwidth}
         \centering
         \begin{annotationimage}{width=1.0\linewidth, trim={3.05cm 2.65cm 0 0}, clip}{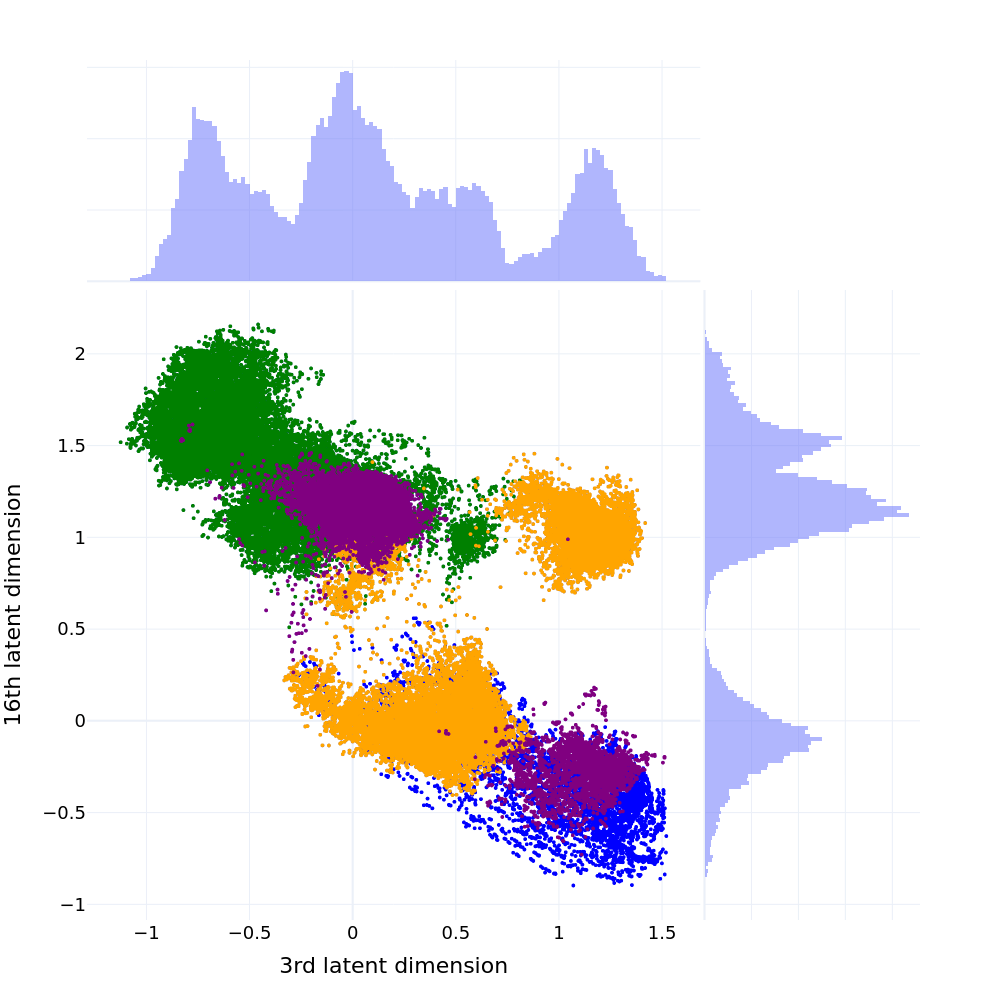}
            \draw[coordinate label = {\normalsize{$z_2$} at (0.35,-0.05)}];
            \draw[coordinate label = {\rotatebox{90}{\normalsize{$z_{15}$}} at (-0.05,0.3)}];
        \end{annotationimage}
         \caption{Better latent space, $l = 16$}
         \label{fig:better_latent}
     \end{subfigure}
     \hfill
     \begin{subfigure}[b]{0.329\textwidth}
         \centering
         \begin{annotationimage}{width=1.0\linewidth, trim={3.05cm 2.65cm 0 0}, clip}{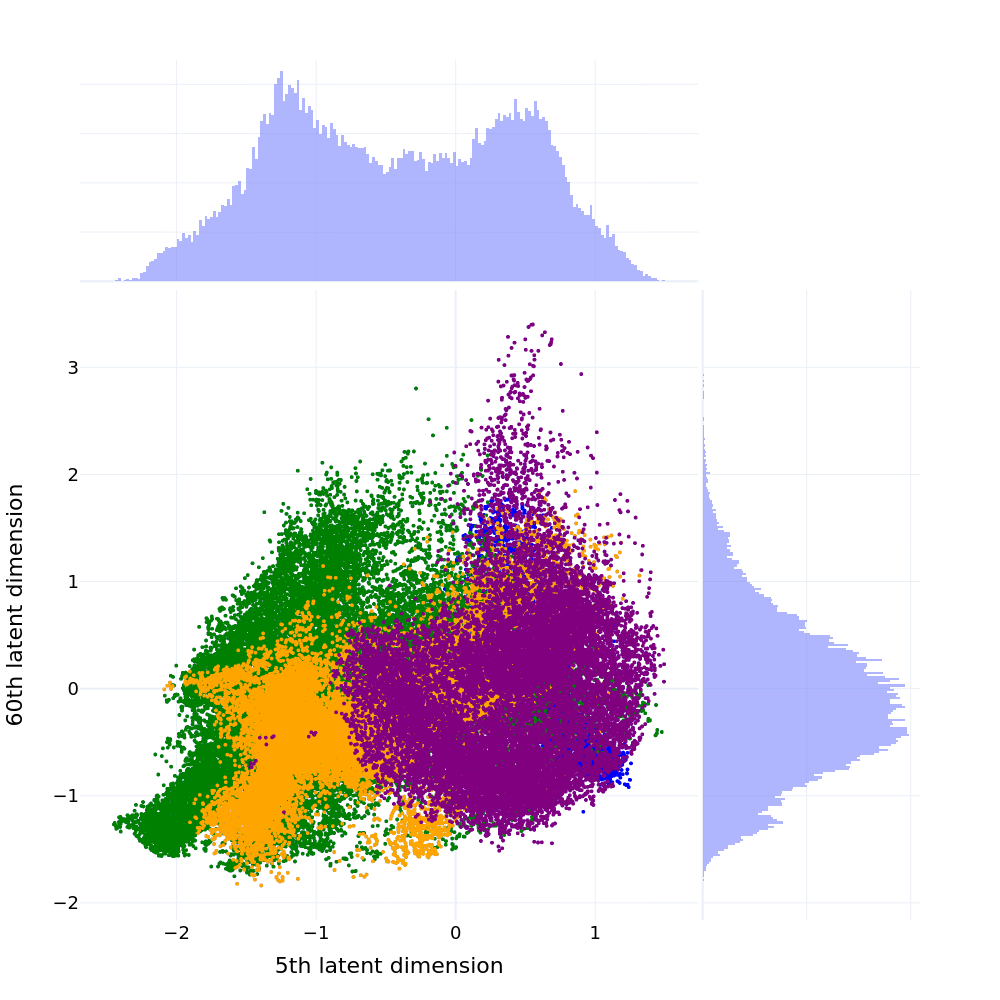}
            \draw[coordinate label = {\normalsize{$z_4$} at (0.35,-0.05)}];
            \draw[coordinate label = {\rotatebox{90}{\normalsize{$z_{59}$}} at (-0.05,0.3)}];
        \end{annotationimage}
         \caption{Good latent space, $l = 64$}
         \label{fig:best_latent}
     \end{subfigure}
    \caption{Example 2D slices of the 32D latent space with well distributed and poorly distributed points and $l = [4, 16,64]$, (blue - O\textsubscript{2}, purple - Ar, green - SF\textsubscript{6} and orange - CF\textsubscript{4}).}
    \label{fig:latent_spaces}
\end{figure*}

\begin{table*}[htbp!]
  \caption{\label{tab:AEresults}Results of autoencoder model training.}
  \footnotesize
  \lineup
  \begin{tabular*}{\textwidth}{@{}l*{15}{@{\extracolsep{0pt plus 12pt}}l}}
\br

\multirow{2}{2cm}{Latent units ($l$)} & \multicolumn{3}{c}{Spectra MSE}                                         & \multicolumn{3}{c}{Image MSE}                                           \\
 & Train    & Validation & Test     &  Train    & Validation & Test \\
4 & $2.09\times10^{-4}$  & $2.09\times10^{-4}$   & $2.06\times10^{-4}$ & $1.06\times10^{-3}$  & $1.10\times10^{-3}$    & $1.09\times10^{-3}$  \\
8 & $4.50\times10^{-5}$ & $4.59\times10^{-5}$   & $4.71\times10^{-5}$ &  $1.03\times10^{-4}$ & $1.04\times10^{-4}$   & $1.07\times10^{-4}$  \\
16 & $2.83\times10^{-5}$ & $2.93\times10^{-5}$   & $2.87\times10^{-5}$ &  $7.64\times10^{-5}$ & $7.56\times10^{-5}$   & $7.77\times10^{-5}$  \\
32 & $1.29\times10^{-5}$ & $1.30\times10^{-5}$   & $1.31\times10^{-5}$ & $4.22\times10^{-5}$ & $4.19\times10^{-5}$   & $4.30\times10^{-5}$  \\
64 & $8.07\times10^{-6}$ & $8.25\times10^{-6}$   & $8.06\times10^{-6}$ & $3.69\times10^{-5}$ & $3.65\times10^{-5}$   & $3.74\times10^{-5}$  \\

\br
\end{tabular*}\end{table*}\normalsize




\subsection{Tool to latent model architecture}

Our decoder model can be used on its own for generative modelling, by randomly sampling over values of $\mathbf{z}$ we can generate random output spectra and images from our model, however, this is of limited practical use. To make this model into a synthetic data generator we need an additional model to learn to map from tool parameters $\textbf{t}$ to the latent space, $\textbf{z} = f(\textbf{t})$. This is similar in its way of thinking to text-to-image models, such as Stable Diffusion \cite{rombachHighResolutionImageSynthesis2022}, where the model is trained with pairs of text descriptions and images. In this work we train an additional model to produce latent representations, $\mathbf{z}$, from tool parameters that match the ones from their associated image and spectra pair. The parameters used were the net power on the ICP coil, table power, gas flows and pressure.

The model is a multi-layer perceptron, a stack of identical dense neural network layers, trained with the latent representations, $\mathbf{z}$, as a supervised objective. As we do not have a reference architecture for this model, and since its small size and low complexity mean it is fast to train, we used KerasTuner \cite{omalleyKerasTuner2019} to carry out a multi-objective Bayesian-optimisation of the number of dense layers, number of neurons and the learning rate for each of models with $l = [4,16,32,64]$. We considered using the top 5 models as an ensemble, but we did not see a discernible improvement. 


\subsection{Evaluating the quality of unsupervised models}

It is inherently difficult to evaluate the quality of unsupervised models as we do not have direct access to the objective that we are optimising for. In this work we trained our models to reduce the MSE between the original image and spectra and their reconstructions. However, this does not tell us if our latent space has useful information, i.e. if the encoding into this space is a useful empirical model of plasma information contained in the diagnostic data and/or if the latent representations produces by our tool model project back to the correct diagnostic information.

To evaluate this we have to create surrogate objectives that we believe provide us some insight into how well we achieve our underlying objective. The simplest method is to look at the performance of our models on our hold-out test data, if the model has simply memorised the input data and cannot generalise and interpolate between the trained data we will see poor reconstructions of the test data. To evaluate if our latent representation is useful for generating synthetic data we can look at the distribution of points in the latent space and make subjective judgements, e.g. large gaps and spaces between points are areas that cannot be sensibly interpolated across by our generative decoder. To evaluate the empirical quality of the models we can evaluate their behaviour around known mode transitions like the E-H mode, comparing trends to previous experimental data and changes in gas stoichiometry.

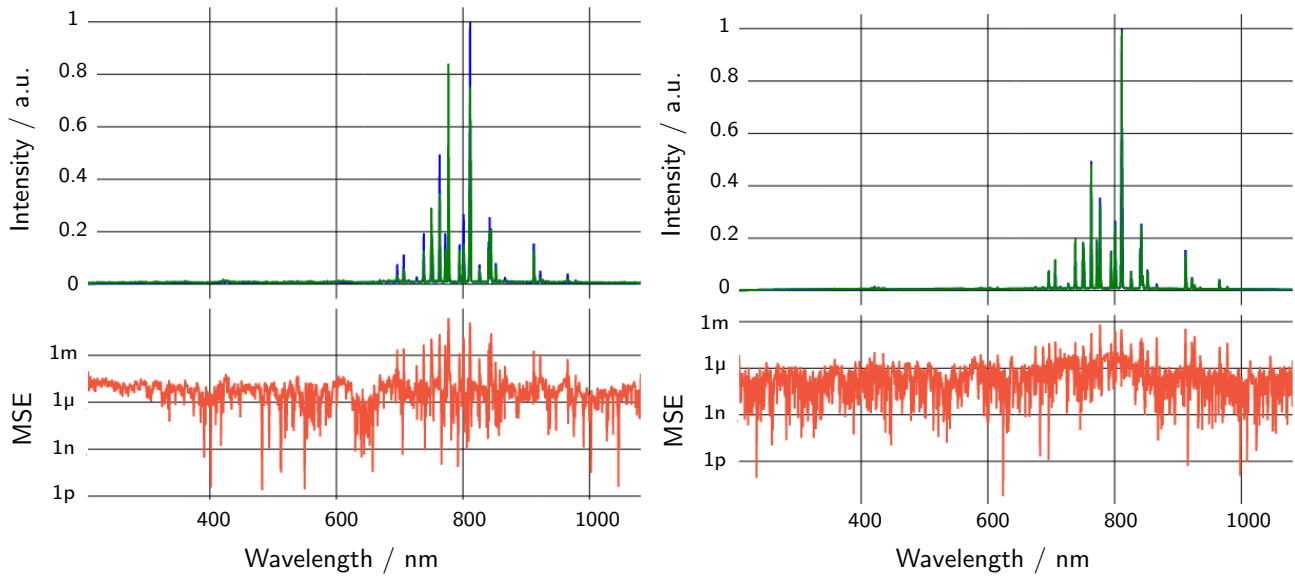
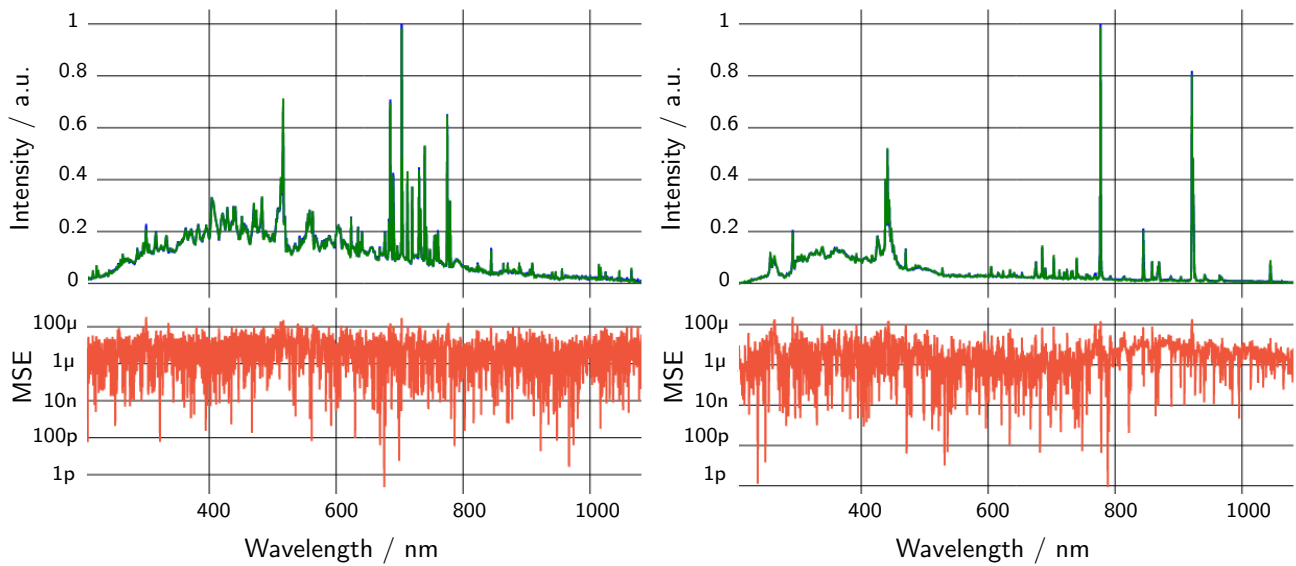
\begin{figure*}[h]
    \centering
    \begin{subfigure}[t]{0.495\textwidth}
         \centering
         \begin{annotationimage}{width=1.0\linewidth, trim={3.3cm 2.5cm 0 5cm}, clip}{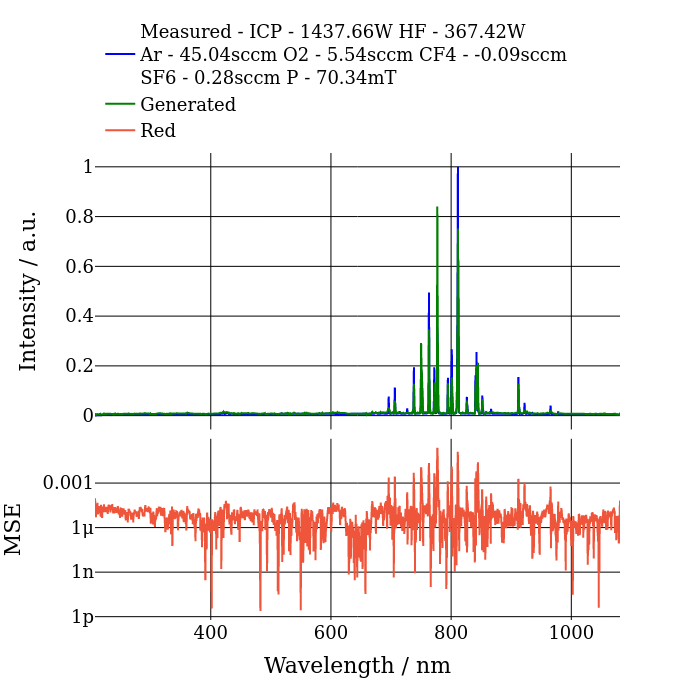}
            \draw[coordinate label = {\footnotesize{400} at (0.2,-0.02)}];
            \draw[coordinate label = {\footnotesize{600} at (0.4,-0.02)}];
            \draw[coordinate label = {\footnotesize{800} at (0.6,-0.02)}];
            \draw[coordinate label = {\footnotesize{1000} at (0.8,-0.02)}];
            \draw[coordinate label = {\normalsize{Wavelength / nm} at (0.4,-0.1)}];
            \draw[coordinate label = {\footnotesize{0} at (-0.02, 0.44)}];
            \draw[coordinate label = {\footnotesize{0.2} at (-0.02, 0.545)}];
            \draw[coordinate label = {\footnotesize{0.4} at (-0.02, 0.65)}];
            \draw[coordinate label = {\footnotesize{0.6} at (-0.02, 0.75)}];
            \draw[coordinate label = {\footnotesize{0.8} at (-0.02, 0.855)}];
            \draw[coordinate label = {\footnotesize{1} at (-0.02, 0.955)}];
            \draw[coordinate label = {\footnotesize{1m} at (-0.035, 0.305)}];
            \draw[coordinate label = {\footnotesize{1\textmu} at (-0.035, 0.21)}];
            \draw[coordinate label = {\footnotesize{1n} at (-0.035, 0.12)}];
            \draw[coordinate label = {\footnotesize{1p} at (-0.035, 0.03)}];
            \draw[coordinate label = {\rotatebox{90}{\normalsize{MSE}} at (-0.1,0.175)}];
            \draw[coordinate label = {\rotatebox{90}{\normalsize{Intensity / a.u.}} at (-0.1,0.7)}];
        \end{annotationimage}
         \caption{$l = 4$, ICP - 2910.9 W, Table - 392.6 W, Ar - 35.3 sccm, pressure - 14.06 mT}
         \label{fig:l4-ar_o2_recon}
     \end{subfigure}
     \hfill
    \begin{subfigure}[t]{0.495\textwidth}
         \centering
         \begin{annotationimage}{width=1.0\linewidth, trim={3.3cm 2.7cm 0 5cm}, clip}{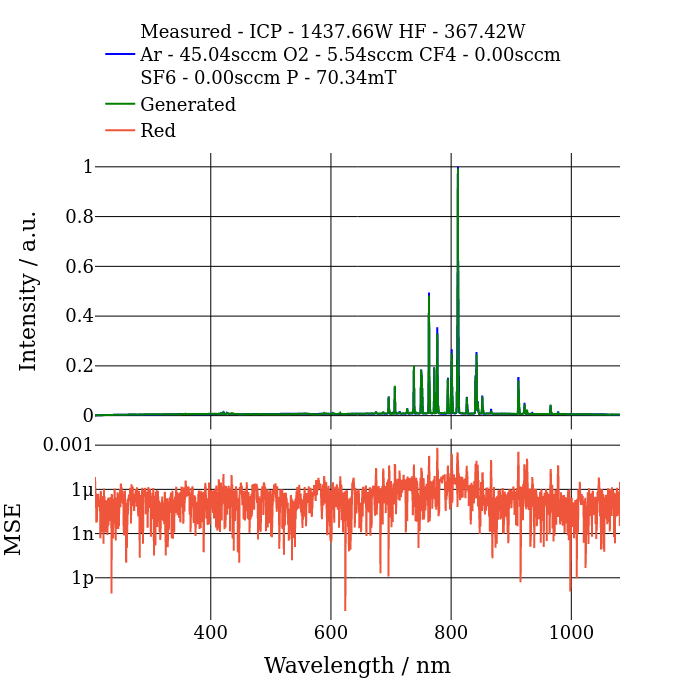}
            \draw[coordinate label = {\footnotesize{400} at (0.2,-0.02)}];
            \draw[coordinate label = {\footnotesize{600} at (0.4,-0.02)}];
            \draw[coordinate label = {\footnotesize{800} at (0.6,-0.02)}];
            \draw[coordinate label = {\footnotesize{1000} at (0.8,-0.02)}];
            \draw[coordinate label = {\normalsize{Wavelength / nm} at (0.4,-0.1)}];
            \draw[coordinate label = {\footnotesize{0} at (-0.02, 0.44)}];
            \draw[coordinate label = {\footnotesize{0.2} at (-0.02, 0.545)}];
            \draw[coordinate label = {\footnotesize{0.4} at (-0.02, 0.65)}];
            \draw[coordinate label = {\footnotesize{0.6} at (-0.02, 0.75)}];
            \draw[coordinate label = {\footnotesize{0.8} at (-0.02, 0.855)}];
            \draw[coordinate label = {\footnotesize{1} at (-0.02, 0.955)}];
            \draw[coordinate label = {\footnotesize{1m} at (-0.035, 0.37)}];
            \draw[coordinate label = {\footnotesize{1\textmu} at (-0.035, 0.285)}];
            \draw[coordinate label = {\footnotesize{1n} at (-0.035, 0.1925)}];
            \draw[coordinate label = {\footnotesize{1p} at (-0.035, 0.095)}];
            \draw[coordinate label = {\rotatebox{90}{\normalsize{MSE}} at (-0.1,0.175)}];
            \draw[coordinate label = {\rotatebox{90}{\normalsize{Intensity / a.u.}} at (-0.1,0.7)}];
        \end{annotationimage}
         \caption{$l = 64$, ICP - 1437.7 W, Table - 367.4 W, Ar - 45 sccm, O\textsubscript{2} - 5.5 sccm, pressure - 70.3 mT}
         \label{fig:l64-ar_o2_recon}
     \end{subfigure}
     \vskip\baselineskip
     \begin{subfigure}[t]{0.495\textwidth}
         \centering
         \begin{annotationimage}{width=1.0\linewidth, trim={3.1cm 2.7cm 0 5cm}, clip}{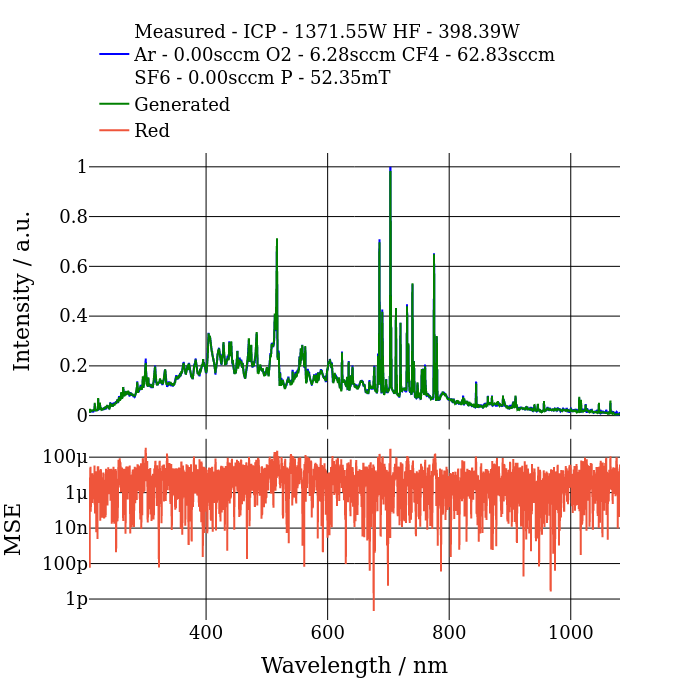}
            \draw[coordinate label = {\footnotesize{400} at (0.2,-0.02)}];
            \draw[coordinate label = {\footnotesize{600} at (0.4,-0.02)}];
            \draw[coordinate label = {\footnotesize{800} at (0.6,-0.02)}];
            \draw[coordinate label = {\footnotesize{1000} at (0.8,-0.02)}];
            \draw[coordinate label = {\normalsize{Wavelength / nm} at (0.4,-0.1)}];
            \draw[coordinate label = {\footnotesize{0} at (-0.02, 0.44)}];
            \draw[coordinate label = {\footnotesize{0.2} at (-0.02, 0.545)}];
            \draw[coordinate label = {\footnotesize{0.4} at (-0.02, 0.65)}];
            \draw[coordinate label = {\footnotesize{0.6} at (-0.02, 0.75)}];
            \draw[coordinate label = {\footnotesize{0.8} at (-0.02, 0.855)}];
            \draw[coordinate label = {\footnotesize{1} at (-0.02, 0.955)}];
            \draw[coordinate label = {\footnotesize{100\textmu} at (-0.048, 0.35)}];
            \draw[coordinate label = {\footnotesize{1\textmu} at (-0.035, 0.275)}];
            \draw[coordinate label = {\footnotesize{10n} at (-0.04, 0.2)}];
            \draw[coordinate label = {\footnotesize{100p} at (-0.048, 0.125)}];
            \draw[coordinate label = {\footnotesize{1p} at (-0.035, 0.05)}];
            \draw[coordinate label = {\rotatebox{90}{\normalsize{MSE}} at (-0.1,0.245)}];
            \draw[coordinate label = {\rotatebox{90}{\normalsize{Intensity / a.u.}} at (-0.1,0.7)}];
        \end{annotationimage}
         \caption{$l = 64$, ICP - 1371.6 W, Table - 398.4 W, O\textsubscript{2} - 6.3 sccm, CF\textsubscript{4} - 62.8 sccm, pressure - 52.4 mT}
         \label{fig:cf4_o2_recon}
     \end{subfigure}
     \hfill
     \begin{subfigure}[t]{0.495\textwidth}
         \centering
         \begin{annotationimage}{width=1.0\linewidth, trim={3.1cm 2.7cm 0 5cm}, clip}{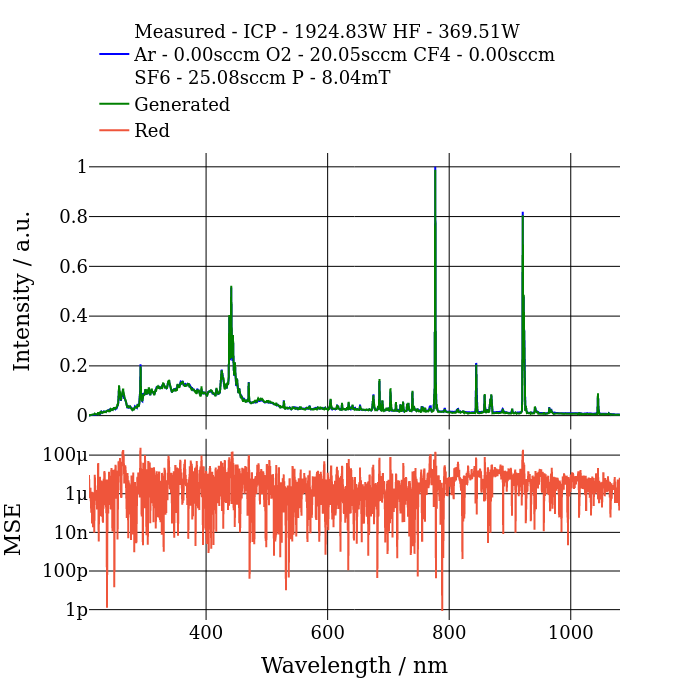}
            \draw[coordinate label = {\footnotesize{400} at (0.2,-0.02)}];
            \draw[coordinate label = {\footnotesize{600} at (0.4,-0.02)}];
            \draw[coordinate label = {\footnotesize{800} at (0.6,-0.02)}];
            \draw[coordinate label = {\footnotesize{1000} at (0.8,-0.02)}];
            \draw[coordinate label = {\normalsize{Wavelength / nm} at (0.4,-0.1)}];
            \draw[coordinate label = {\footnotesize{0} at (-0.02, 0.44)}];
            \draw[coordinate label = {\footnotesize{0.2} at (-0.02, 0.545)}];
            \draw[coordinate label = {\footnotesize{0.4} at (-0.02, 0.65)}];
            \draw[coordinate label = {\footnotesize{0.6} at (-0.02, 0.75)}];
            \draw[coordinate label = {\footnotesize{0.8} at (-0.02, 0.855)}];
            \draw[coordinate label = {\footnotesize{1} at (-0.02, 0.955)}];
            \draw[coordinate label = {\footnotesize{100\textmu} at (-0.048, 0.35)}];
            \draw[coordinate label = {\footnotesize{1\textmu} at (-0.035, 0.275)}];
            \draw[coordinate label = {\footnotesize{10n} at (-0.04, 0.2)}];
            \draw[coordinate label = {\footnotesize{100p} at (-0.048, 0.125)}];
            \draw[coordinate label = {\footnotesize{1p} at (-0.035, 0.05)}];
            \draw[coordinate label = {\rotatebox{90}{\normalsize{MSE}} at (-0.1,0.245)}];
            \draw[coordinate label = {\rotatebox{90}{\normalsize{Intensity / a.u.}} at (-0.1,0.7)}];
        \end{annotationimage}
         \caption{$l = 64$, ICP - 1924.8 W, Table - 369.5 W, O\textsubscript{2} - 20.1 sccm, SF\textsubscript{6} - 25.1 sccm, pressure - 8 mT}
         \label{fig:sf6_o2_recon}
     \end{subfigure}
    \caption{Measured OES and reconstructions in Ar/O\textsubscript{2} for $l = 4 $ and $64$, CF\textsubscript{4}/O\textsubscript{2}, and SF\textsubscript{6}/O\textsubscript{2} plasmas, green line is the measured spectrum, blue line is the reconstructed spectrum. Given the difficulty of telling them apart, the red line below shows the mean squared error at each wavelength.}
    \label{fig:oes_recons}
\end{figure*}

\section{Properties of the latent space}\label{latentspaces}

The overall aim of latent space modelling is to project input data onto a manifold in the latent space while preserving information and relationships within the data that are physically real and sensible, whilst not overfitting on spurious relationships that are not physically real or sensible. To make our latent representation usable we would like it to have some properties, for points to be close to a normal distribution, for points that are close in the real space (i.e. two plasmas that are similar to each other) to be close in the latent space and the reverse to be true, and for the latent space to be interpolatable, i.e. we can smoothly move through the latent space from one area to another without sharp discontinuities. 

\setlength{\tabcolsep}{0pt}

\begin{figure*}
    \centering
    
        \begin{tabular}{cccc}
             \includegraphics[]{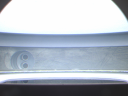}
            & \includegraphics[]{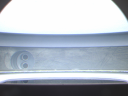}
            & \includegraphics[]{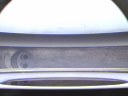}
            & \includegraphics[]{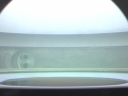}\\[-4pt]
             \includegraphics[]{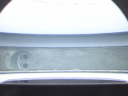}
            & \includegraphics[]{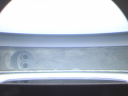}
            & \includegraphics[]{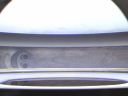}
            & \includegraphics[]{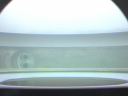}\\[-4pt]
        \end{tabular}%
    \caption{Measured images and reconstructions in Ar/O\textsubscript{2} ($l = 4$, $l = 64$), CF\textsubscript{4}/O\textsubscript{2}, and SF\textsubscript{6}/O\textsubscript{2} plasmas. Top row is original images, bottom is reconstructions.}
    \label{fig:picturegrid}
\end{figure*}

Many of these properties can be gained by simply using a large enough deep learning model with enough data. Large neural networks are inherently self-regularising \cite{barrettImplicitGradientRegularization2020} and with increasing size, reach a point where their outputs become Lipschitz continuous \cite{bubeckUniversalLawRobustness2021}. When training generative models on existing benchmark datasets, it is possible to use measures of image similarity to evaluate the performance of the model, such as the Fréchet inception distance \cite{heuselGANsTrainedTwo2017}. However, these use pre-trained image classification networks to evaluate the quality of generated images. If our data was similar to the data used to train the classification network these methods can be used, or if you have some labelled data you can fine-tune one of these models for this use case. However, an OES of an Argon plasma has little similarity to images of planes and cats (which are typically employed in pre-trained networks) so we would not have any guarantee that these methods would work. This is an area of active research in the field of generative modelling and so in time new evaluation methods may appear that overcome this issue.

Without a quantitative measure of performance we are left with qualitative evaluations of our generative capabilities. The simplest is to look at the distribution of points in the latent space. If our model and dataset are large enough and the model is well trained, our latent space should be well behaved -- close to a normal distribution and interpolatable. In figure \ref{fig:latent_spaces} we show three examples of the latent space of a trained model, `bad', `better' and `good'. The bad example shows a latent space that is extremely sparse and has significant spikes in the concentration of points, it would be very difficult to interpolate between points in this space as it has significant discontinuities and no meaningful representation moving off the central axis the points are stretched across. In the better example most of the points are reasonably close, although we have a strongly multimodal distribution and has separated into two clusters that would be extremely difficult to interpolate between. The good representation shows what we are looking for, our points are more smoothly distributed and there are no discontinuities within the latent space itself.

Unfortunately we cannot always expect our data to be perfectly well behaved like our `good' representation. We cannot rely on the assumption that our data is independent and identically distributed. The conditions of one plasma are affected by the history of plasmas within that tool and we expect our latent space to encode some physically real multi-modal distributions, like E-H mode transitions, different gas stoichiometries and pressure regimes. Figure \ref{fig:best_latent} shows a `good' representation, the latent space is smooth and interpolatable, but one dimension has a bimodal distribution. We expect to see different physical modes in the data form independent normal distributions in the latent space and as long as it is physically possible to transition between these modes, and we have data covering the mode transition, the latent space can be used to interpolate between these modes.

\section{Evaluating the generative model}\label{geneval}

A summary of the results from training the autoencoder model is given in table \ref{tab:AEresults}. The training data split was used for directly training each model, the validation split was used to independently evaluate model performance for hyperparamter optimisation of the model learning rate. The optimal hyperparameters found for the training and fine-tuning step are summarised in table \ref{tab:trainingsetttings}. The test split was kept as a holdout set for final model evaluation and was not used at any time during training and hyperparameter optimisation. The test and train errors are very close for all latent space sizes, indicating that the model has not overfit to the training data. In Figures \ref{fig:oes_recons} and \ref{fig:picturegrid} we show 3 random examples, from the test split, $l = 64$, of the original and reconstructed data in each of our three gas mixtures and $l = 4$ for the Ar/O\textsubscript{2} example. The error on the reconstruction is extremely low for $l = 64$, but as can be seen in table \ref{tab:AEresults} and figure \ref{fig:l4-ar_o2_recon}, the reconstruction error decreases significantly for larger latent space size. In particular, figure \ref{fig:l4-ar_o2_recon} shows that the small latent space model makes significant errors in reconstructing the relative height of peaks in the spectrum and at $l = 64$ these are greatly minimised.


\begin{figure}[t]  
    \centering
    \begin{subfigure}[t]{0.495\textwidth}
        \centering
        \includegraphics[width=\textwidth]{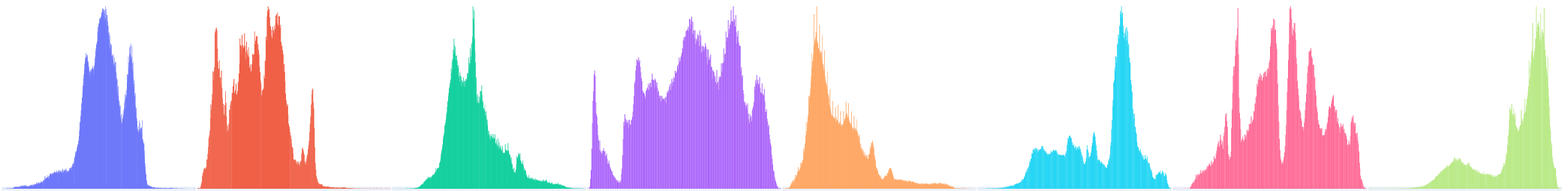}
        \caption{$l = 8$}
         \label{fig:z8_hist}
     \end{subfigure}
    \vskip\baselineskip
    \begin{subfigure}[t]{0.495\textwidth}
        \centering
        \includegraphics[width=\textwidth]{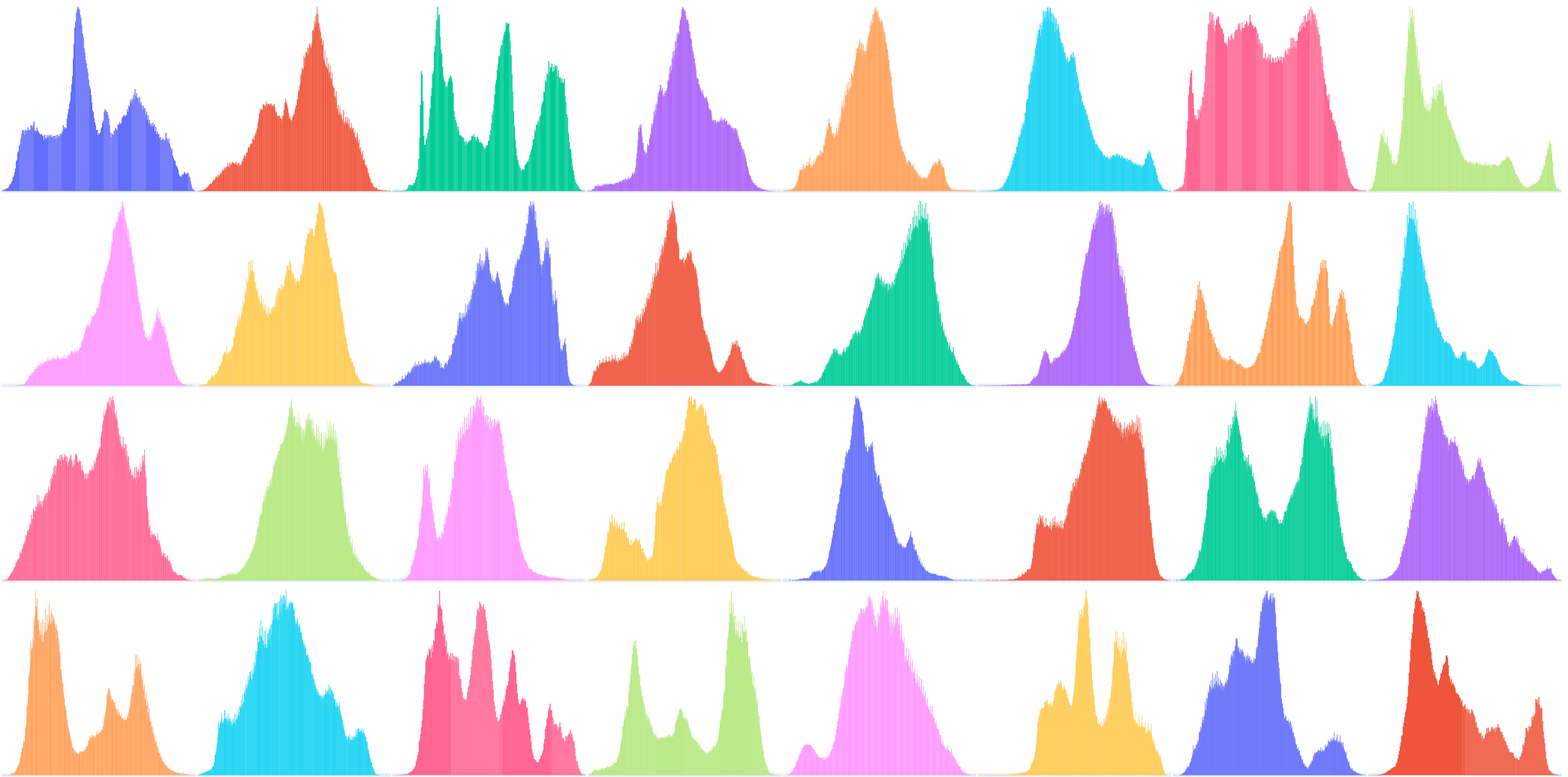}
        \caption{$l = 32$}
         \label{fig:z32_hist}
     \end{subfigure}
    \vskip\baselineskip
    \begin{subfigure}[t]{0.495\textwidth}
        \centering
        \includegraphics[width=\textwidth]{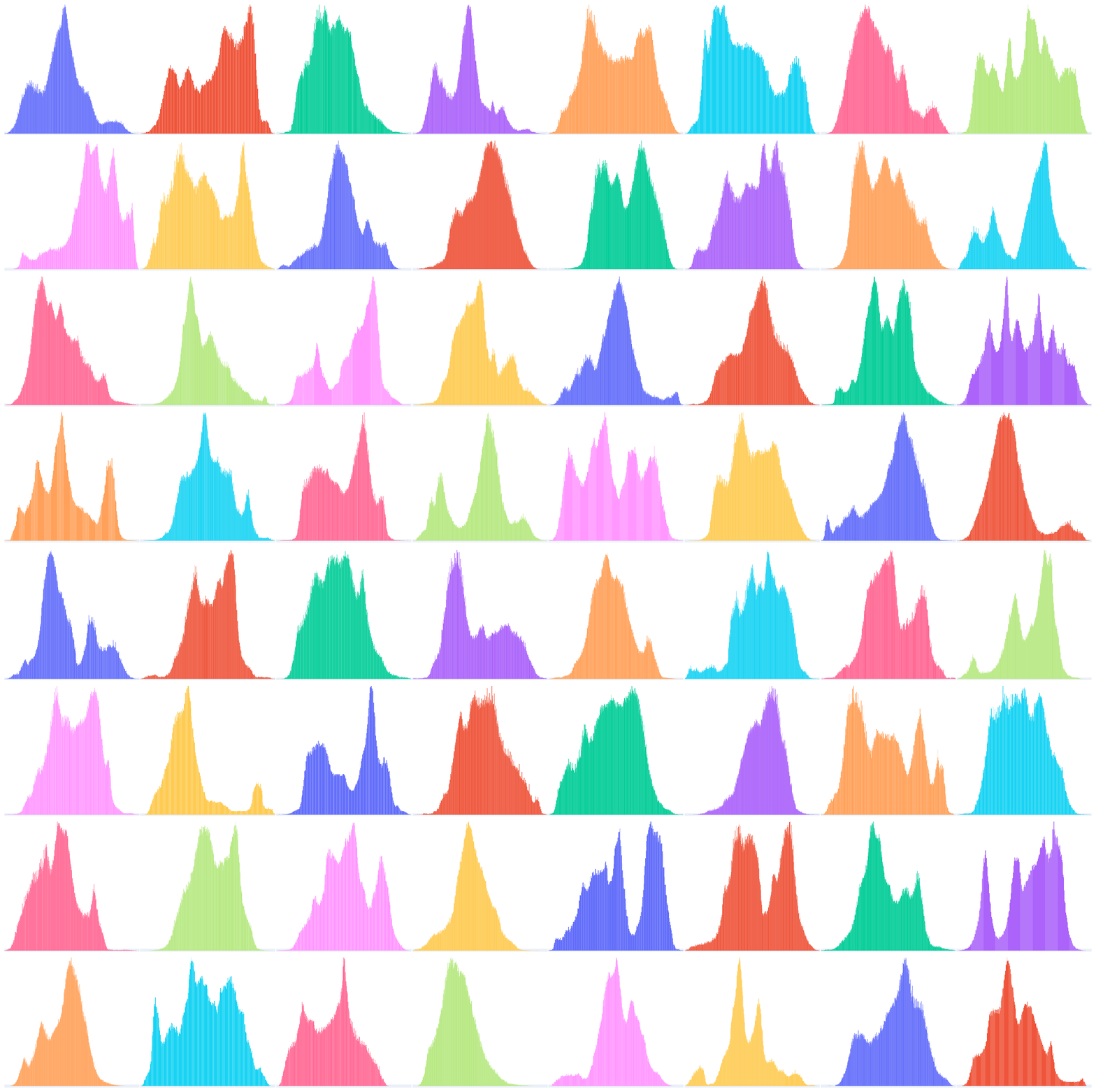}
        \caption{$l = 64$}
         \label{fig:z64_hist}
     \end{subfigure}
    \caption{Histograms of the distribution of points in each latent dimension space for all image spectra pairs in the test set.}
    \label{fig:latent_hist_grid}
\end{figure}

\begin{figure}[t]
    \centering
    \begin{subfigure}[t]{0.495\textwidth}
         \centering
         \begin{annotationimage}{width=0.98\linewidth, trim={2.5cm 2.5cm 0 4cm}, clip}{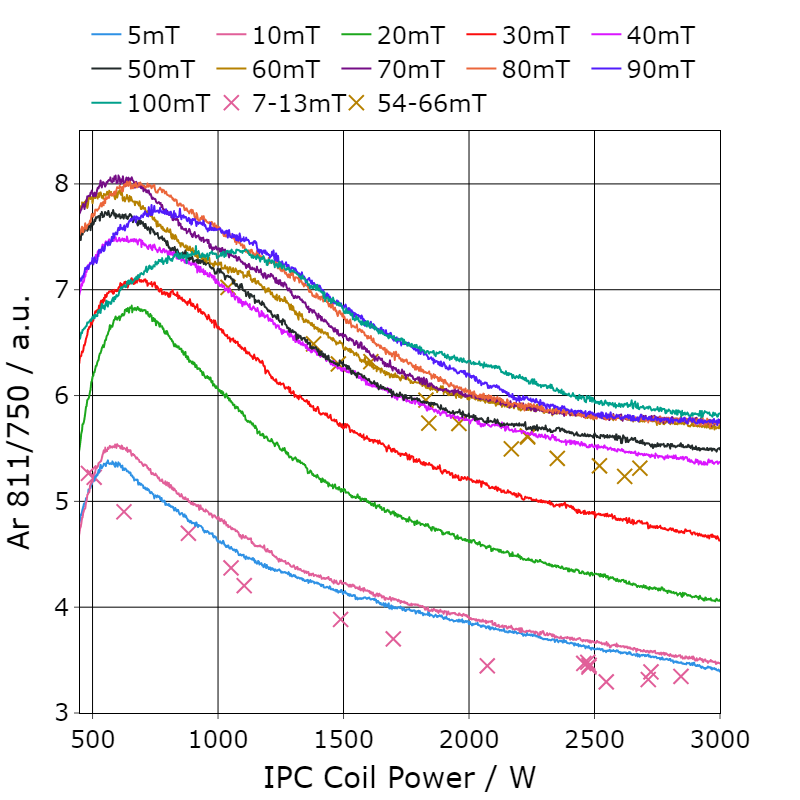}
            \draw[coordinate label = {\footnotesize{500} at (0.035,-0.02)}];
            \draw[coordinate label = {\footnotesize{1000} at (0.2,-0.02)}];
            \draw[coordinate label = {\footnotesize{1500} at (0.375,-0.02)}];
            \draw[coordinate label = {\footnotesize{2000} at (0.55,-0.02)}];
            \draw[coordinate label = {\footnotesize{2500} at (0.715,-0.02)}];
            \draw[coordinate label = {\footnotesize{3000} at (0.8875,-0.02)}];
            \draw[coordinate label = {\normalsize{ICP coil power / W} at (0.45,-0.1)}];
            \draw[coordinate label = {\footnotesize{3} at (-0.02, 0.03)}];
            \draw[coordinate label = {\footnotesize{4} at (-0.02, 0.2)}];
            \draw[coordinate label = {\footnotesize{5} at (-0.02, 0.37)}];
            \draw[coordinate label = {\footnotesize{6} at (-0.02, 0.54)}];
            \draw[coordinate label = {\footnotesize{7} at (-0.02, 0.71)}];
            \draw[coordinate label = {\footnotesize{8} at (-0.02, 0.89)}];
            \draw[coordinate label = {\rotatebox{90}{\normalsize{$I_{811.5}/I_{750.5}$ / a.u.}} at (-0.075,0.5)}];
        \end{annotationimage}
         \caption{Ar 811.5 nm / 750.4 nm line ratio}
         \label{fig:ICP_sweep_ar_811_750}
     \end{subfigure}
     \hfill
    \begin{subfigure}[t]{0.495\textwidth}
         \centering
         \begin{annotationimage}{width=0.98\linewidth, trim={2.5cm 2.5cm 0 4cm}, clip}{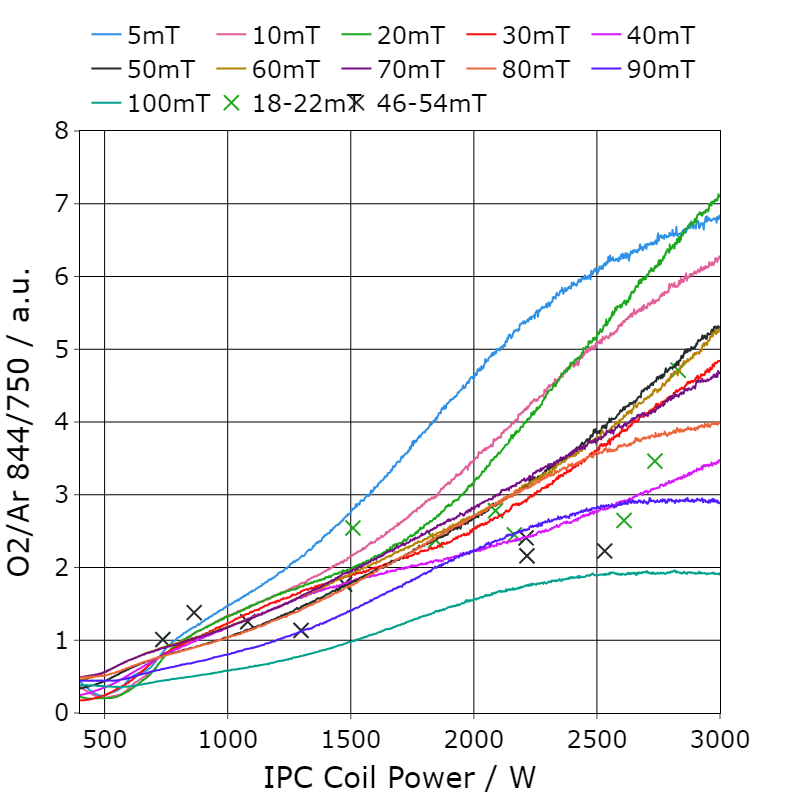}
            \draw[coordinate label = {\footnotesize{500} at (0.05,-0.02)}];
            \draw[coordinate label = {\footnotesize{1000} at (0.22,-0.02)}];
            \draw[coordinate label = {\footnotesize{1500} at (0.385,-0.02)}];
            \draw[coordinate label = {\footnotesize{2000} at (0.55,-0.02)}];
            \draw[coordinate label = {\footnotesize{2500} at (0.725,-0.02)}];
            \draw[coordinate label = {\footnotesize{3000} at (0.8875,-0.02)}];
            \draw[coordinate label = {\normalsize{ICP coil power / W} at (0.4,-0.1)}];
            \draw[coordinate label = {\footnotesize{0} at (-0.02, 0.03)}];
            \draw[coordinate label = {\footnotesize{1} at (-0.02, 0.15)}];
            \draw[coordinate label = {\footnotesize{2} at (-0.02, 0.265)}];
            \draw[coordinate label = {\footnotesize{3} at (-0.02, 0.38)}];
            \draw[coordinate label = {\footnotesize{4} at (-0.02, 0.5)}];
            \draw[coordinate label = {\footnotesize{5} at (-0.02, 0.62)}];
            \draw[coordinate label = {\footnotesize{6} at (-0.02, 0.735)}];
            \draw[coordinate label = {\footnotesize{7} at (-0.02, 0.855)}];
            \draw[coordinate label = {\footnotesize{8} at (-0.02, 0.97)}];
            \draw[coordinate label = {\rotatebox{90}{\normalsize{$I_{844.6}/I_{750.5}$ / a.u.}} at (-0.075,0.5)}];
        \end{annotationimage}
         \caption{O\textsubscript{2} 844.6 nm / Ar 750.4 nm line ratio}
         \label{fig:ICP_sweep_ar_o2_844_750}
     \end{subfigure}
    \caption{Sweep across from 400-3000 W ICP power in Argon and Ar/O\textsubscript{2}, lines are generated data and crosses are measured points with 10\% of the swept values. Sweeps repeated at \textcolor{pxdark0}{5 mT}, \textcolor{pxdark1}{10 mT}, \textcolor{pxdark2}{20 mT}, \textcolor{pxdark3}{30 mT}, \textcolor{pxdark4}{40 mT}, \textcolor{pxdark5}{50 mT}, \textcolor{pxdark6}{60 mT}, \textcolor{pxdark7}{70 mT}, \textcolor{pxdark8}{80 mT}, \textcolor{pxdark9}{90 mT} and \textcolor{pxdark10}{100 mT}}
    \label{fig:ICP_sweep_ar_o2}
\end{figure}
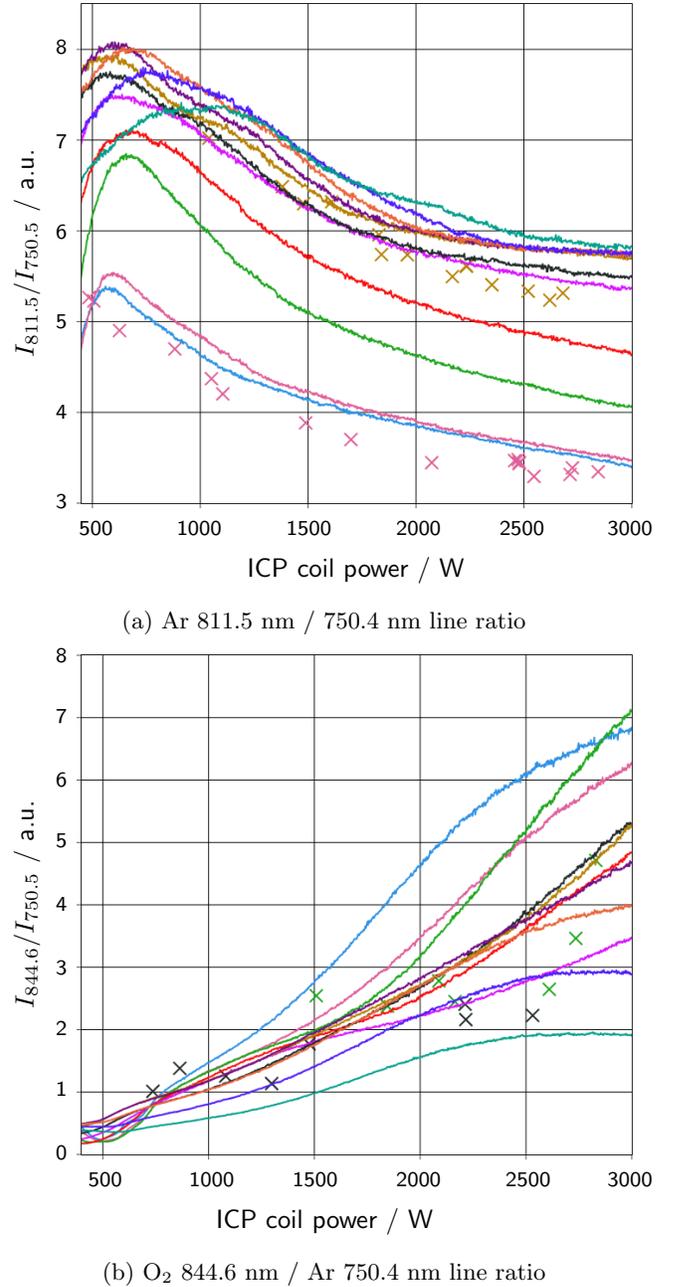

\begin{figure*}[t]
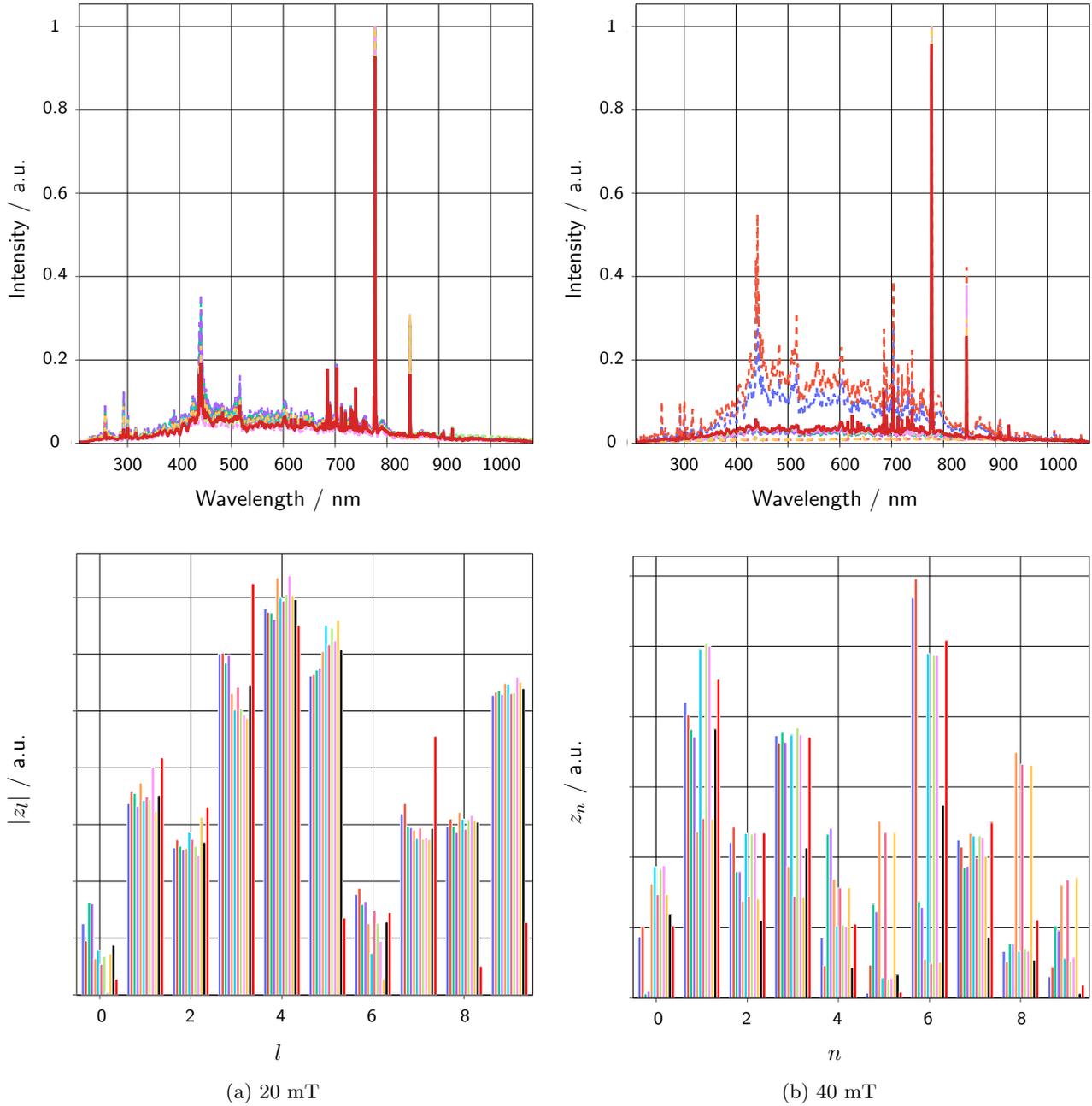

    \centering
    \begin{subfigure}[t]{0.495\textwidth}
         \centering
         \begin{annotationimage}{width=0.95\linewidth, trim={1.8cm 1.7cm 0 2.6cm}, clip}{graphs/nearest_records_2400_300_0_10_10_0_20_l64}
            \draw[coordinate label = {\footnotesize{300} at (0.11,-0.02)}];
            \draw[coordinate label = {\footnotesize{400} at (0.215,-0.02)}];
            \draw[coordinate label = {\footnotesize{500} at (0.315,-0.02)}];
            \draw[coordinate label = {\footnotesize{600} at (0.42,-0.02)}];
            \draw[coordinate label = {\footnotesize{700} at (0.52,-0.02)}];
            \draw[coordinate label = {\footnotesize{800} at (0.625,-0.02)}];
            \draw[coordinate label = {\footnotesize{900} at (0.725,-0.02)}];
            \draw[coordinate label = {\footnotesize{1000} at (0.83,-0.02)}];
            \draw[coordinate label = {\normalsize{Wavelength / nm} at (0.4,-0.1)}];
            \draw[coordinate label = {\footnotesize{0} at (-0.03, 0.03)}];
            \draw[coordinate label = {\footnotesize{0.2} at (-0.03, 0.215)}];
            \draw[coordinate label = {\footnotesize{0.4} at (-0.03, 0.4)}];
            \draw[coordinate label = {\footnotesize{0.6} at (-0.03, 0.58)}];
            \draw[coordinate label = {\footnotesize{0.8} at (-0.03, 0.77)}];
            \draw[coordinate label = {\footnotesize{1} at (-0.03, 0.95)}];
            \draw[coordinate label = {\rotatebox{90}{\normalsize{Intensity / a.u.}} at (-0.1,0.5)}];
        \end{annotationimage}
         \label{fig:cf4_sweep_10_10_o2}
     \end{subfigure}
     \hfill
    \begin{subfigure}[t]{0.495\textwidth}
         \centering
         \begin{annotationimage}{width=0.95\linewidth, trim={1.8cm 1.7cm 0 2.6cm}, clip}{graphs/nearest_records_2400_300_0_10_10_0_40_l64}
            \draw[coordinate label = {\footnotesize{300} at (0.11,-0.02)}];
            \draw[coordinate label = {\footnotesize{400} at (0.215,-0.02)}];
            \draw[coordinate label = {\footnotesize{500} at (0.315,-0.02)}];
            \draw[coordinate label = {\footnotesize{600} at (0.42,-0.02)}];
            \draw[coordinate label = {\footnotesize{700} at (0.52,-0.02)}];
            \draw[coordinate label = {\footnotesize{800} at (0.625,-0.02)}];
            \draw[coordinate label = {\footnotesize{900} at (0.725,-0.02)}];
            \draw[coordinate label = {\footnotesize{1000} at (0.83,-0.02)}];
            \draw[coordinate label = {\normalsize{Wavelength / nm} at (0.4,-0.1)}];
            \draw[coordinate label = {\footnotesize{0} at (-0.03, 0.03)}];
            \draw[coordinate label = {\footnotesize{0.2} at (-0.03, 0.215)}];
            \draw[coordinate label = {\footnotesize{0.4} at (-0.03, 0.4)}];
            \draw[coordinate label = {\footnotesize{0.6} at (-0.03, 0.58)}];
            \draw[coordinate label = {\footnotesize{0.8} at (-0.03, 0.77)}];
            \draw[coordinate label = {\footnotesize{1} at (-0.03, 0.95)}];
            \draw[coordinate label = {\rotatebox{90}{\normalsize{Intensity / a.u.}} at (-0.1,0.5)}];
        \end{annotationimage}
         \label{fig:cf4_sweep_40_10_o2}
     \end{subfigure}
     \vskip\baselineskip
     \vspace{-10pt}
     \begin{subfigure}[t]{0.495\textwidth}
         \centering
         \begin{annotationimage}{width=0.95\linewidth, trim={1.9cm 1.7cm 0 2.6cm}, clip}{graphs/nearest_latents_2400_300_0_10_10_0_20_l64}
            \draw[coordinate label = {\footnotesize{0} at (0.06,-0.02)}];
            \draw[coordinate label = {\footnotesize{2} at (0.23,-0.02)}];
            \draw[coordinate label = {\footnotesize{4} at (0.405,-0.02)}];
            \draw[coordinate label = {\footnotesize{6} at (0.58,-0.02)}];
            \draw[coordinate label = {\footnotesize{8} at (0.76,-0.02)}];
            \draw[coordinate label = {\normalsize{$l$} at (0.4,-0.1)}];
            \draw[coordinate label = {\rotatebox{90}{\normalsize{$\lvert z_l\rvert$ / a.u.}} at (-0.1,0.5)}];
        \end{annotationimage}
         \caption{20 mT}
         \label{fig:cf4_sweep_10_10_o2_latent}
     \end{subfigure}
     \hfill
    \begin{subfigure}[t]{0.495\textwidth}
         \centering
         \begin{annotationimage}{width=0.95\linewidth, trim={1.9cm 1.7cm 0 2.6cm}, clip}{graphs/nearest_latents_2400_300_0_10_10_0_40_l64}
            \draw[coordinate label = {\footnotesize{0} at (0.06,-0.02)}];
            \draw[coordinate label = {\footnotesize{2} at (0.23,-0.02)}];
            \draw[coordinate label = {\footnotesize{4} at (0.405,-0.02)}];
            \draw[coordinate label = {\footnotesize{6} at (0.58,-0.02)}];
            \draw[coordinate label = {\footnotesize{8} at (0.76,-0.02)}];
            \draw[coordinate label = {\normalsize{$n$} at (0.4,-0.1)}];
            \draw[coordinate label = {\rotatebox{90}{\normalsize{$z_n$ / a.u.}} at (-0.1,0.5)}];
        \end{annotationimage}
         \caption{40 mT}
         \label{fig:cf4_sweep_40_10_o2_latent}
     \end{subfigure}
    \caption{Generated spectra and first 10 latent coordinates at 2400 W ICP, 300 W Table, 10 sccm O\textsubscript{2}, 10 sccm CF\textsubscript{4} and spectra of 10 nearest points in the dataset. In the upper plots, the solid line is the generated spectra and the dotted lines are the nearest 10. In the lower plots the first 10 bars are the latent coordinates of the 10 nearest, the black line is their average and the red is the generated latent.}
    \label{fig:cf4_sweep_10_o2}
\end{figure*}

To evaluate the quality of our model's latent space we can look at the distribution of points encoded into the latent space. In figure \ref{fig:latent_hist_grid} we can see the type of distributions we have in our latent space for $l = 8$ and $64$. We can make a qualitative assessment of the quality of the latent space for generative modelling. For $l = 8$ the distributions show some sections that are smoothly and normally distributed, but has a large number of discontinuities (spikes and troughs) and are all strongly multimodal. For $l = 32$ and $64$ some of our latent dimensions have a uni-modal distribution, but the majority have multi-modal distributions, and there is some complexity in the distributions.  There are spikes present in $l = 32$ suggesting that some mode collapse has occurred (e.g. multiple measurements mapped to the exact same place in the latent space), but not $l = 64$. In $l = 64$ there are no gaps in the latent space, although there are areas of very low density of points between parts of the distribution in a few of the latent dimensions, but $l = 32$ does have two areas of nearly zero density, suggesting a gap in the latent space.. These qualitative assessments suggest that our $l = 64$ model can be used for generative modelling as we can smoothly interpolate between different areas of the latent without discontinuities, but the smaller $l = 8$ is unsuitable and $l = 32$ would be suitable for most areas, but would struggle around its discontinuities.

\section{Results of synthetic experiments}\label{synexp}

To carry out a synthetic experiment we use our tool-to-latent model, $\textbf{z} = f(\textbf{t})$, to produce latent representations, $\mathbf{z}$, and our two decoder branches, $\mathbf{i} = g(\mathbf{z})$,  $\mathbf{s} = h(\mathbf{z})$, to generate spectra and images. We can generate an image spectra pair for one experiment point in 0.13 s/0.79 s on GPU/CPU, can compute a batch of 128 points in 0.25 s/51.22 s and a batch of 1024 in 1.34 s on an A100 GPU. In the simplest form, we can generate the expected spectra and image at a desired set of powers, pressures and gas mixture. We can also simply perform more complex experiments where we sweep across parameters in fine steps very quickly. Figure \ref{fig:ICP_sweep_ar_o2} shows a simple experiment where we sweep from 400-3000 W applied to the ICP source, 1024 steps, in pure argon and 8 sccm Ar, 50 sccm O\textsubscript{2} at at different 11 pressures from 5-100 mT. We plot the line ratio of the Ar 811.5 nm and 750.4 nm lines in pure Ar and the ratio of the O\textsubscript{2} 844.6 nm and Ar 750.4 nm lines in the Ar/O\textsubscript{2} mixture.

In figure \ref{fig:ICP_sweep_ar_811_750} we show the variation in $(I_{811.5}/I_{750.5})$ ratio with power at pressures between 5 and 100 mT, at 10 and 60 mT we also plot the ratio at points in the data set that are close to the sweep. We can see that the points in the data are reasonably close to the generated data and follow the same trend. The overall trend in the data is in agreement with other experimental data by Czerwiec and Graves \cite{czerwiecModeTransitionsLow2004}, although their reactor was a significantly different geometry. The trend in power shows a linear rise to the E-H mode transition point around 500-600 W and then decreases. Their data is at higher pressures, above 100 mT, and shows no change with pressure, our model shows a strong trend in an increase in $(I_{811.5}/I_{750.5})$ from 10-40 mT, then showing similar behaviour with little change with increasing pressure.

In figure \ref{fig:ICP_sweep_ar_o2_844_750} we show the variation in $(I_{844.6}/I_{750.5})$ ratio with power at pressures between 5 and 100 mT, at 20 and 50 mT we also plot the ratio at points in the data set that are close to the sweep. The points in the data show general agreement with the trends in the data, but the scatter in the points is quite high. The overall trend in the $(I_{844.6}/I_{750.5})$ ratio is in good agreement with earlier work by Fuller et al. \cite{fullerCharacterizationTransformerCoupled2000} with a relatively linear rise with applied power.

\section{Limitations of the model and future work}\label{limit}

The encoder model is able to embed any image / spectra pair into the latent space and very accurately decode them back into the real measurement space. Differences between the real plasma conditions of these measurements are represented by different coordinates in the latent space. When using the encoder model to monitor a plasma, the latent space representation will capture dynamic changes in the plasma over time. However, our tool to latent model is very simplistic, it can only map a set of powers, gas flows and pressures to their average coordinate in the latent space, it cannot capture any dynamics. 

We show an example of this in figure \ref{fig:cf4_sweep_10_o2}, at two pressures in a CF\textsubscript{4}/O\textsubscript{2} plasma, we show the spectra generated at the latent coordinate produced by the tool to latent model and the 10 nearest spectra to this point in the dataset. At 40 mT there is a high variation in the spectra around this area as each point will have had a different history and will each be at different points of rising or falling power in the data collection sweep. This is reflected in the latent representations of these different plasmas, but our tool encoder finds a latent representation that produces an average of these spectra. At 20 mT, there is much less variation in both the measured spectra and latent representation and so there is close agreement between all measurements and generated spectra.

To overcome this issue we do not need to make any modifications to the autoencoder model itself, the latent space representation is capable of representing changes in the plasma and does not collapse to a single point for similar plasmas. We would need to replace our simple tool-to-latent model with a more complex model to account for trajectory of powers and pressures in the experiment. This could be achieved with a sequence-to-sequence model, where the sequence of output latent representations is able to account for previous conditions. This represents one of the advantages of this approach, the unsupervised learning approach allows us to easily disaggregate different parts of a problem and combine the parts of our autoencoder with different models to achieve different goals and these models can be trained with different data sources, where data much more limited or measurements more difficult.


\section{Open source release of the dataset, trained models and code}\label{foss}


The underlying dataset is available at \url{https://doi.org/10.5281/zenodo.7704879}, configured as the train/validation/test splits used in the paper and is released under the Creative Commons Attribution 4.0 International. The model code and trained models are available \href{https://github.com/gregdaly/generative_modelling_for_optical_plasma_diagnostics}{here} and are released under the MIT license. An example notebook of using the model is available \href{https://colab.research.google.com/github/gregdaly/generative_modelling_for_optical_plasma_diagnostics/blob/master/generative_decoder_demo.ipynb}{here} and is released under the MIT license.

\section{Conclusion}
\label{conc}

We have demonstrated that recent advances in generative modelling can be applied to optical diagnostics in low-temperature plasmas. These approaches require a heavily automated approach to experiments, to allow large amounts of data to be gathered in a reasonable amount of time. Large autoencoder models can be trained, using existing open source libraries and model architectures, for a low cost on cloud GPUs or in a relatively short time on local GPU clusters. 

We have shown that the latent space of autoencoders, trained on real plasma diagnostic data, is very sensitive to the size of the latent space. Any implicit bias to produce a model with the smallest number of parameters must be balanced by ensuring that the latent space is smooth and interpolatable if we want the model to be useful or have any capacity for generalisation.

Once trained, these autoencoders provide a low-cost method to generate large volumes of synthetic data for use in other work, such as validating or creating models. This is achieved by training an additional model to sample the latent space in the way required for the synthetic experiment. We have demonstrated this capability with a simple model to map tool inputs into the latent space and generate synthetic data that shows good agreement with experimental data in Argon and Ar/O\textsubscript{2} plasmas.

Large autoencoders can become a foundational building block for a wide array of plasma physics experiments and models when trained with large datasets of simple, but information dense diagnostics. The encoder can produce latent representations of diagnostics that are smoothly interpolatable and sensibly separates similar and dissimilar plasmas. These latent representations can be used for monitoring experiments or as inputs for other predictive models. The decoder can produce realistic and accurate data from latent representations and can be extended with auxiliary models to make a powerful generative model for synthetic experiments, which we aim to exploit in future work.

\ack


This work was supported by the Engineering and Physical Research Council (grant number EP/L016389/1) and Oxford Instruments Plasma Technology (OIPT). The authors would like to thank Dr Thomas Miller and Mr Kris Balo at OIPT for their assistance in modifying the plasma processing tool for these experiments.  We would also like to thank Dr Frazer Anderson at OIPT for giving us access to the labs and GPU cluster of OIPT's Innovation Development Group for this work.

\section*{References}

\bibliographystyle{iopart-num.bst}
\bibliography{main.bib}

\end{document}